\documentclass{aa}  

\usepackage{graphicx}
\usepackage{grffile}
\usepackage[varg]{txfonts}
\usepackage{subfigure}
\usepackage{amssymb}
\usepackage{cancel}
\usepackage{bm}
\usepackage[draft]{hyperref}
\usepackage{color}
\usepackage{footnote}
\usepackage{eqnarray}
\usepackage{verbatim}
\begin{document}

   \title{Inferring giant planets from ALMA millimeter continuum\\ and line observations in (transition) disks}
   
\titlerunning{Inferring giant planets from mm continuum\\ and line observations}

   \author{S. Facchini\inst{1}   
           \and
           P. Pinilla\inst{2}\thanks{Hubble Fellow}
           \and
           E.~F. van Dishoeck\inst{1,}\inst{3}
           \and
           M. de Juan Ovelar\inst{4}
          }

   \institute{Max-Planck-Institut f\"ur Extraterrestrische Physik, Giessenbachstrasse 1, 85748 Garching, Germany,\\\email{facchini@mpe.mpg.de}    
                         \and
                 Department of Astronomy, University of Arizona, 933 North Cherry Avenue, Tucson, AZ, United States.
                 \and
                 Leiden Observatory, Leiden University, P.O. Box 9513, NL-2300 RA Leiden, The Netherlands               \and
                 Astrophysics Research Institute, Liverpool John Moores University, 146 Brownlow Hill, Liverpool L3 5RF, UK
             }

   \date{Received; accepted}

 
  \abstract
   {Radial gaps or cavities in the continuum emission in the IR-mm wavelength
range are potential signatures of protoplanets embedded in their natal protoplanetary disk are . Hitherto, models have relied on the combination of mm continuum observations and near-infrared (NIR) scattered light images to put constraints on the properties of embedded planets. Atacama Large Millimeter/submillimeter Array  (ALMA) observations are now probing spatially resolved rotational line emission of CO and other chemical species. These observations can provide complementary information on the mechanism carving the gaps in dust and additional constraints on the purported planet mass.}
   {We investigate whether the combination of ALMA continuum and CO line observations can constrain the presence and mass of planets embedded in protoplanetary disks.}
   {We post-processed azimuthally averaged 2D hydrodynamical simulations of planet-disk models, in which the dust densities and grain size distributions are computed with a dust evolution code that considers radial drift, fragmentation, and growth. The simulations explored various planet masses ($ 1\,M_{\rm J}\leq M_{\rm p}\leq15\,M_{\rm J}$) and turbulent parameters ($ 10^{-4}\leq \alpha\leq10^{-3}$). The outputs were then post-processed with the thermochemical code DALI, accounting for the radially and vertically varying dust properties. We obtained the gas and dust temperature structures, chemical abundances, and synthetic emission maps of both thermal continuum and CO rotational lines. This is the first study combining hydrodynamical simulations, dust evolution, full radiative transfer, and chemistry to predict gas emission of disks hosting massive planets.}
   {All radial intensity profiles of $^{12}$CO, $^{13}$CO, and C$^{18}$O show a gap at the planet location. The ratio between the location of the gap as seen in CO and the peak in the mm continuum at the pressure maximum outside the orbit of the planet shows a clear dependence on planet mass and is independent of disk viscosity for the parameters explored in this paper. Because of the low dust density in the gaps, the dust and gas components can become thermally decoupled and the gas becomes colder than the dust. The gaps seen in CO are due to a combination of gas temperature dropping at the location of the planet and of the underlying surface density profile. Both effects need to be taken into account and disentangled when inferring gas surface densities from observed CO intensity profiles; otherwise, the gas surface density drop at the planet location can easily be overestimated. CO line ratios across the gap are able to quantify the gas temperature drop in the gaps in observed systems. Finally, a CO cavity not observed in any of the models, only CO gaps, indicating that one single massive planet is not able to explain the CO cavities observed in transition disks, at least without additional physical or chemical mechanisms.}
   {}

   \maketitle
%
\section{Introduction}

\label{sec:intro}

Protoplanetary disks consist of gaseous and dusty material orbiting in nearly Keplerian motion around a newborn star. Soon after their discovery, these disks were indicated as the birthplace of planets;  this hypothesis has been strengthened by the tentative detection of planet candidates still embedded in their natal environment from imaging and radial velocity techniques \citep{kraus_12,2014ApJ...792L..23R,2017arXiv171011393R,2014ApJ...792L..22B,2013ApJ...766L...1Q,2015ApJ...807...64Q,2015Natur.527..342S,2016ApJ...826..206J}. However, these direct (or indirect) imaging techniques are extremely challenging, thus other methods of determining the presence and masses of embedded planets are needed.

The high sensitivity and angular resolution provided by the Acatama Large Millimeter/submillimeter Array (ALMA) and by near-infrared (NIR) instruments as SPHERE/VLT \citep{2008SPIE.7014E..18B} and GPI/Gemini \citep{2008SPIE.7015E..18M} provide complementary information on the potential presence of protoplanets in protoplanetary disks by looking at the details of the disk structures, which can be affected by the interaction with the embedded planet(s) when the planet is massive enough. Two main effects are expected when a protoplanet is gravitationally interacting with its parent disk: spirals are launched by the planet and gaps are opened in the dust (and possibly gas) surface density. In this paper we focus only on the gap signature.

Hydrodynamical simulations of planet-disk interactions have been used to predict the emission in both scattered light and thermal continuum of disks hosting a planet \citep[e.g.,][]{2006MNRAS.373.1619R,pinilla_12,2012A&A...547A..58G,2013A&A...549A..97R}. In particular, hydrodynamical simulations also
provide quantitative observational diagnostics to determine the mass of embedded planets from high angular resolution observations of ring-like structures in disks \citep[e.g.,][]{2013A&A...560A.111D,2015ApJ...806L..15K,2016ApJ...818..158A,2016MNRAS.459.2790R,2017MNRAS.469.1932D,2017ApJ...835..146D}. Almost all of these simulations rely on the  combination of scattered light and (sub-)mm continuum observations. In particular, semi-analytic relations have been derived for some properties of spatially resolved observations as a function of planet mass, disk turbulence, and aspect ratio, for example, for the ratio of the cavity radii in transition disks as seen in scattered light and (sub-)mm continuum \citep{2013A&A...560A.111D}, the gap depth in (sub-)mm continuum \citep{2015ApJ...806L..15K}, the gap width as traced by scattered light images \citep{2016MNRAS.459.2790R,2017ApJ...835..146D}, and the distance of the (sub-)mm ring from the gap seen in scattered light observations \citep{2016MNRAS.459.2790R}. Many other studies have focused on the feasibility of detecting rings and gaps in scattered light and thermal continuum emission, where the rings and gaps may be formed by a plethora of physical mechanisms that do not require the presence of planets \citep[e.g.,][]{2012MNRAS.419.1701R,2015ApJ...813L..14B,2015A&A...574A..68F,2016A&A...590A..17R}.

In this paper, we explore how the combination of (sub-)mm continuum and gas line observations can be used to infer the potential presence and mass of massive planets embedded in their natal protoplanetary disk. In particular, we aim to study thoroughly the effects of the perturbation driven by such planets in the spatially resolved gas emission; in  particular we focus on CO rotational lines that are readily observed together with the mm continuum.

Very few studies have focused on the observational signatures of the gas component across the gap carved by a planet. So far, two different approaches have been used. The first considers a parametrized density structure, where gas  and dust temperatures and chemistry are computed on top of that. For example, \citet{2015ApJ...807....2C} computed molecular emission from a disk hosting a gap, considering the additional heating source of an embedded planet and focusing on the azimuthal asymmetry caused by this second source in the system. \citet{isella_16,2017A&A...600A..72F,2017ApJ...835..228T} also considered a parametrized disk density structure with gaps and rings to reproduce spatially resolved molecular observations in specific sources. The second approach has been to post-process 3D hydrodynamical simulations of planet-disk interaction leading to the formation of a gap in the disk structure \citep{2015A&A...579A.105O}. In this case, the dust distribution is assumed to follow perfectly the gas distribution in both density and temperature, and no chemical calculations are performed.

In this paper, we relax most assumptions taken by both approaches. In order to obtain reliable gas emission maps, temperature estimates of both dust and gas components are needed together with chemical abundances. Second, a physically motivated density structure of both gas and dust is needed, since the two do not necessarily follow each other. \citet{2017A&A...605A..16F} have shown that the vertical and radial variations of the dust grain size distribution in disks can significantly affect the gas temperature and CO emission via both thermal and chemical effects. Gaps and cavities carved by massive planets are expected to suffer these effects significantly. In this paper, we thus post-process hydrodynamical simulations of disks hosting one massive planet ($M_{\rm p}\geq1\,M_{\rm J}$) using the method presented in \citet{2017A&A...605A..16F}. In particular, We use the thermochemical code DALI \citep[Dust And LInes;][]{2012A&A...541A..91B,2013A&A...559A..46B} to post-process the hydrodynamical simulations of planet-disk interaction presented in \citet{2016MNRAS.459L..85D}, where dust evolution is computed on top of the gas hydrodynamics. The DALI code self-consistently computes the gas temperature and chemical abundances, which can then provide synthetic emission maps of gas lines.

The massive planets considered in this work can lead to the formation of transition disks with large mm cavities. Recently, ALMA has allowed us to image the gas component of a few large transition disks within the (sub-)mm continuum cavity \citep{2013Sci...340.1199V,2014A&A...562A..26B,2015ApJ...798...85P,2015A&A...579A.106V,2016A&A...585A..58V,2017ApJ...836..201D,2017A&A...600A..72F}. In all cases, except SR24S and SR21 \citep{2015A&A...579A.106V,2016A&A...585A..58V,2017ApJ...839...99P}, a depletion of gas surface density, as probed by a CO cavity, is observed within the dust cavity. By the word cavity we mean a significant depletion of CO within the inner dust (sub-mm) ring and no increase toward the star, as opposed to a gap. These CO cavities are usually associated with a depletion in the gas surface density in the inner regions of transition disks caused by one or multiple planets \citep[e.g.,][]{zhu_11,2012ApJ...755....6Z,pinilla_12,2015ApJ...809...93D,2015A&A...573A...9P,2017ApJ...839..100F}. However, this picture is somehow in friction with the high accretion rates, $\dot{M}> 10^{-9}-10^{-8}\,M_\odot$\,yr$^{-1}$, measured for most of these transition disks with large cavities \citep[e.g.,][]{2014A&A...568A..18M,2014prpl.conf..497E}. If the gas surface densities are indeed as low as traced by the low CO emission inside the dust cavities because gas is cleared by multiple planets \citep[e.g.][]{zhu_11,2017ApJ...839..100F}, such high mass accretion rates cannot be sustained for long. We note that this problem arises no matter what is causing the gas depletion, unless material is rapidly infalling onto the star with radial supersonic velocities from the outer disk \citep{2014ApJ...782...62R,2016PASA...33....5O,2017arXiv170104627Z}. Indeed some observations are now tracing hints of non-Keplerian motions within the central (sub-)mm cavity \citep{2012ApJ...757..129R,2015ApJ...811...92C,2017A&A...597A..32V,2017ApJ...840...23L}, but higher angular and spectral resolution observations are still needed to test this scenario \citep[see][]{2016PASA...33...13C} and distinguish it from warped inner disks \citep[e.g.,][]{2017MNRAS.466.4053J,2018MNRAS.473.4459F}. In this paper, we explore whether the CO cavities observed in most large transitions disks might be explained by a single massive planet.

The paper is organized as follows: In Section~\ref{sec:methods} we describe the hydrodynamical simulations, dust model, and thermochemical calculations setup used to predict the gas temperature and emission. In Section~\ref{sec:results} we present our results, and in Section~\ref{sec:discussion} we generalize the results to obtain semi-analytical prescriptions to derive planet masses combining CO and (sub-)mm continuum observations. In Section~\ref{sec:concl} we summarize the results and draw our conclusions.

\section{Method and setup}
\label{sec:methods}

We combined the results of the hydrodynamical simulations of planet-disk interaction by \citet{2016MNRAS.459L..85D} with the thermochemical code DALI \citep[][]{2012A&A...541A..91B,2013A&A...559A..46B}. In particular, we computed the 2D (radial and azimuthal) gas density structure using FARGO 2D \citep{2000A&AS..141..165M} and calculated the dust distribution over the azimuthally averaged gas structure using the 1D (radial) dust evolution code by \citet{2010A&A...513A..79B}. We then computed the dust and gas vertical density and temperature with DALI, in which the dust treatment follows the updates by \citet{2017A&A...605A..16F}. The synthetic images for both gas and dust were produced using the DALI ray tracer. The scattered light images shown here are taken from \citet{2016MNRAS.459L..85D}. The details of the method and setup parameters are described below.

\begin{figure}
\center
\includegraphics[width=\columnwidth]{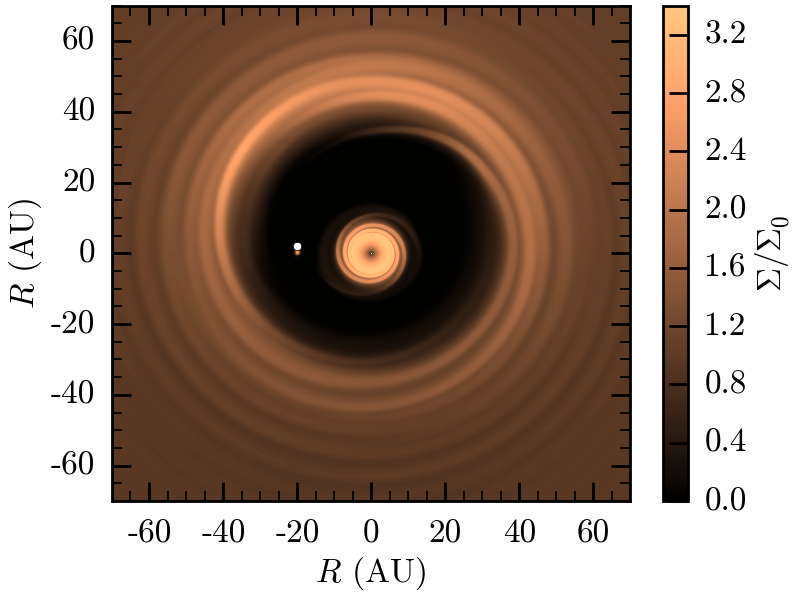}
\caption{Snapshot of one of the hydro-simulations from \citet{2016MNRAS.459L..85D} after $1000$ planet orbits, with $\alpha=10^{-3}$ and $M_{\rm p}=5\,M_{\rm J}$.}
\label{fig:hydro}
\end{figure}

\subsection{Gas hydrodynamics}

The modeling approach for the gas hydrodynamical simulations was described in \citet{pinilla_12,2013A&A...560A.111D}. The 2D gas simulations are run for 1000 orbits at the planet location, which is set at 20\,AU in circular orbit in all simulations (see Fig.~\ref{fig:hydro}). The radial grid is logarithmically sampled with 512 bins between $0.5$ and $140$\,AU with an azimuthal sampling of 1024 points. The simulations assume a power-law temperature profile; the derived aspect ratio $H/R$ is expressed as:

\begin{equation}
H/R=0.05\left(\frac{R}{20\,{\rm AU}}\right)^{1/4},
\label{eq:hoverr}
\end{equation}
where $H$ is the scale height of the disk and $R$ the cylindrical radius. The initial gas surface density is set to $\Sigma_{\rm gas}(R)\propto R^{-1}$. The disk is assigned a mass $M_{\rm disk}=0.0525\,M_\odot$ and the central star $M_*=1\,M_\odot$ in all simulations. We explored simulations with four different planet masses, namely $M_{\rm p}=[1,\ 5,\ 9,\ 15]M_{\rm J}$, where $M_{\rm J}$ is the mass of Jupiter. Finally, the simulations assume a kinematic viscosity $\nu=\alpha H^2 \Omega$ \citep{shakura73}, where $\Omega$ is the orbital frequency, and $\alpha$ is the dimensionless parameter to parametrize the disk turbulence. The simulations consider two values of disk turbulence, where $\alpha=[10^{-3},10^{-4}]$, which agrees with the recent turbulence constraints by \citet{2015ApJ...813...99F,2016A&A...592A..49T}. The accretion onto the planet and planet migration are not considered in these simulations. As already mentioned, our hydro simulations are 2D. \citet{2015ApJ...813...88Z} showed that within the planet mass range explored in this paper the density structures in the midplane of the disk are similar between 2D and 3D simulations. These authors showed that density structure can differ between the two at high altitudes in the disk because of the vertical temperature gradient modifying the local sound speed \citep[see also, e.g.,][]{2017arXiv171103559J}. Further modeling would be required to assess whether these differences at high altitudes can have an impact on the molecular gas emission calculated in this paper.

We do not consider potential variations of the disk scale height that can arise from, for example, inclined planets \citep[e.g.,][]{2017MNRAS.468.4610C}. 

\subsection{Dust radial grain size distribution}
\label{sec:dust}

In order to reconstruct the radial distribution of dust particles, we used the 1D dust evolution code by \citet{2010A&A...513A..79B}, in which grain growth and fragmentation are computed considering radial drift, turbulent motion, and gas drag. The code is run over the azimuthally averaged gas surface density and radial velocities; the 180 size bins range between $1\,\mu$m and $200\,$cm with logarithmic sampling. The gas properties are taken after 1000 planet orbits, when the disk is assumed to be in quasi-steady state \citep[see the Appendix in][for more details about the assumptions of the models]{2016MNRAS.459L..85D}. The dust evolution code computes the grain size distribution at every radial point for an evolution time of $1\,$Myr. We assumed a fragmentation velocity of $10\,$m\,s$^{-1}$ and a solid density of the dust particles of $1.2$\,g\,cm$^{-3}$. Particles colliding at relative speeds lower than $80\%$ of the fragmentation velocity are assumed to stick perfectly, whereas particles colliding between $80$ and $100\%$ of the fragmentation velocity undergo erosion and fragmentation. We did not consider any effect of the radiation pressure caused by the accretion onto the planet, which might affect the dynamics of the small grains, acting as a radial dam for particle sizes $\lesssim1\,\mu$m \citep{2014ApJ...789...59O}.

\subsection{Radial and vertical structures, temperatures, and abundances}

We then used the output of the 2D gas hydrodynamical simulation and 1D dust evolution routine to compute the thermal structure of both gas and dust and the chemical abundances via the code DALI. The method that we used  to import the 1D radial profiles of both gas and dust densities into DALI is similar to that in \citet{2017A&A...605A..16F}. The main limitation of the method used in this paper is that the gas surface density is azimuthally averaged when computing the grain size distribution and dust densities (Section~\ref{sec:dust}) and in the thermochemical post-processing. It is known that very massive planets ($M_{\rm p}\gtrsim 5 M_{\rm J}$) can form some azimuthal asymmetries in the disk gas density structure, as shown in Fig.~\ref{fig:hydro}. These asymmetries are expected to be even more significant in the dust density structures for these extreme cases \citep[e.g.,][]{2017MNRAS.464.1449R}, in particular for large grains. In this paper the azimuthal asymmetries are not modeled.

The code DALI is 2D, where the two spatial dimensions are radius $R$ and height $z$. The radial grid is defined differently from that used in the hydrodynamical simulations. A different grid is needed at this stage because for both the radiative transfer and thermochemistry the inner radius is taken at $0.13\,$AU, defined as the sublimation radius $R_{\rm subl}$ for a star of $3L_\odot$, where the sublimation radius is computed as $R_{\rm subl}=0.07\sqrt{L_*/L_\odot}\,$AU \citep{2001ApJ...560..957D}. A posteriori, we found that the temperature at the inner edge of the disk from our simulations is $\sim1100\,$K, thus slightly lower than the dust sublimation temperature. However, in the calculation of the dust temperature, no viscous heating is considered, which can be important for the determination of the exact location of the sublimation radius \citep[][]{2002apa..book.....F}. Since the exact location of the sublimation radius does not affect the results shown in the paper, which focuses on the outer disk, we just used the simple prescription reported above. The outer radius is still set to 140\,AU. The radial grid is composed of 60 bins between the inner radius and $5\,$AU, sampled logarithmically. Then, a linear grid of 70 bins between $5$ and $40$\,AU is used to sample every $0.5$\,AU in the region where the planet opens a gap. The gap and the radial gradient of gas and dust densities at the edge of the gap are well resolved by the sub-AU grid cells in all simulations. Finally, we used a logarithmic grid between $40$ and $140$\,AU, composed of 30 bins. In order to check numerical convergence, we ran a model with doubled resolution within the gap opened by the planet (i.e., between $5$ and $40$\,AU) and doubled resolution along the vertical direction. A difference $<3\%$ was found in the spatially resolved radial emission profiles of the $^{12}$CO $J$=2-1, 3-2 and 6-5 lines, well below the uncertainties of the chemical model.

The gas and dust surface densities are remapped into the new radial grid by linearly interpolating the hydrodynamical output. Moreover, the same profiles are extrapolated down to the sublimation radius from the inner radius of the hydrodynamical simulations. We tried two different extrapolation schemes, which lead to undistinguishable results for the science addressed in this paper. The first scheme extrapolates the gas surface density by fitting the gas surface density between $1$ and $1.5\,$AU with a power law and then extending the gas surface density down to the sublimation radius with the power-law index derived in the fitting. The second scheme more simply assumes that the gas surface density scales with $R^{-1}$ between the sublimation radius and $1\,$AU. We note that in the hydrodynamical simulation we expected the gas surface density to scale with $R^{-1}$ at large enough radii not to be affected by the inner boundary condition, and at small enough radii not to be modified by the planet torques. From viscous evolution theory, the inner regions of the disk  reaches a quasi-steady state on short timescales, with $\Sigma_{\rm gas}\propto R^{-\gamma}$, where $\gamma$ is defined via the dependence of the kinematic viscosity on radius, i.e., $\nu\propto R^{\gamma}$. The aspect ratio defined with Eq.~\ref{eq:hoverr} implies that $\nu\propto R$, and thus $\Sigma_{\rm gas}\propto R^{-1}$. All the results shown in this manuscript use the first extrapolation scheme.

\begin{figure*}
\begin{center}
\includegraphics[width=\textwidth]{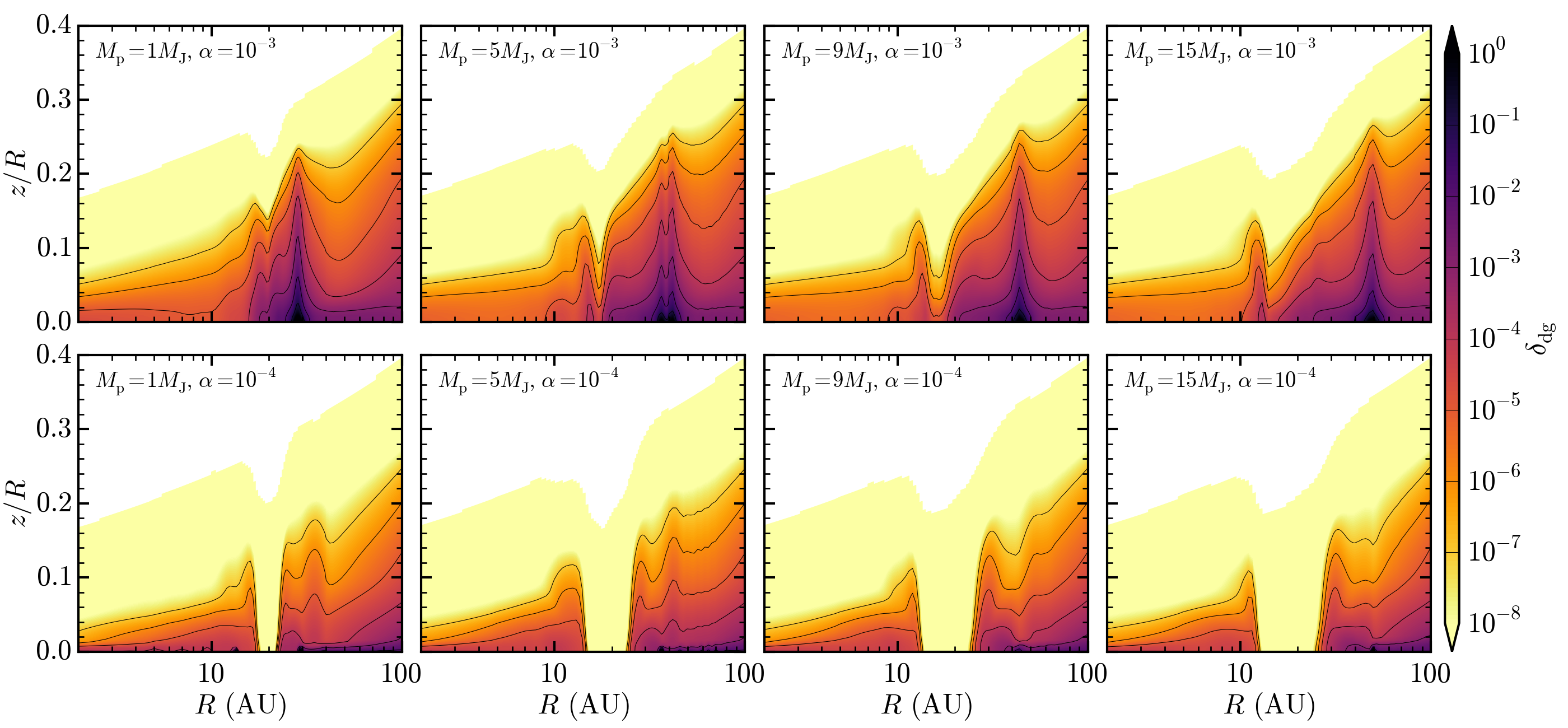}
\end{center}
\caption{Dust-to-gas ratio $\delta_{\rm dg}$ for all models. Top and bottom panels represent $\alpha=10^{-3}$ and $10^{-4}$, respectively. From left to right, $M_{\rm p}=[1,\ 5,\ 9,\ 15]M_{\rm J}$.}
\label{fig:d2g_ratio}
\end{figure*}

The dust particles size grid is extended down to $50\,\AA$ with a total number of logarithmically spaced size bins of 250 between $50\,\AA$ and $200\,$cm. The very small particles are very important both in the continuum radiative transfer and thermochemistry, since they provide the opacities at far-ultraviolet (FUV) wavelengths. The extrapolation is based on the smallest particles available from the 1D dust evolution routine $\Sigma_{\rm dust}(R,a<1\,\mu{\rm m})=\Sigma_{\rm dust}(R,1\,\mu{\rm m})$, where $\Sigma_{\rm dust}(R,a)$ is defined as the dust mass surface density at a radius $R$, with particle sizes between $a$ and $a + da$ (in radius). We thus implicitly set the grain size distribution per unit surface for particles sizes $a<1\,\mu$m to follow $f(a,R)\propto a^{-3}$. The typical power-law exponent observed in the diffusive interstellar medium (ISM) is $q=3.5$ \citep{1977ApJ...217..425M}, but we used a lower value following observations indicating a more top heavy distribution in protoplanetary disks \citep[e.g.,][]{2010A&A...521A..66R,2010A&A...512A..15R,2014prpl.conf..339T}. 

The radial density structure is then expanded in the vertical direction. The gas mass density $\rho_{\rm gas}$ is determined assuming hydrostatic equilibrium in the vertical direction,

\begin{equation}
\rho_{\rm gas} (R,z)= \frac{\Sigma_{\rm gas}(R)}{ \sqrt{2\pi} H} \exp { \left[ -\frac{1}{2} \left( \frac{z}{H} \right)^2 \right] },
\label{eq:gas_dens}
\end{equation}
where $H$ follows the radial dependence of Eq.~\ref{eq:hoverr}. The dust vertical structure is then computed solving the advection-diffusion equation of vertical settling in steady state, as described in Section 2.3 of \citet{2017A&A...605A..16F}. The average grain size at every point of the disk is obtained from the same method of Section 2.4 of \citet{2017A&A...605A..16F} \citep[following][]{2011ApJ...727...76V}. The dust temperatures are calculated with the opacities from Section 2.5 of \citet{2017A&A...605A..16F}, using the continuum DALI radiative transfer. In particular, dust opacities are computed by mass averaging the opacities of every grain size bin at any location in the disk, where for every grain size the mass extinction coefficients are calculated from Mie theory with the \verb!miex! code \citep{2004CoPhC.162..113W}. Optical constants are from \citet{2003ApJ...598.1017D} for graphite and \citet{2001ApJ...548..296W} for silicates.

\begin{figure*}
\begin{center}
\includegraphics[width=\textwidth]{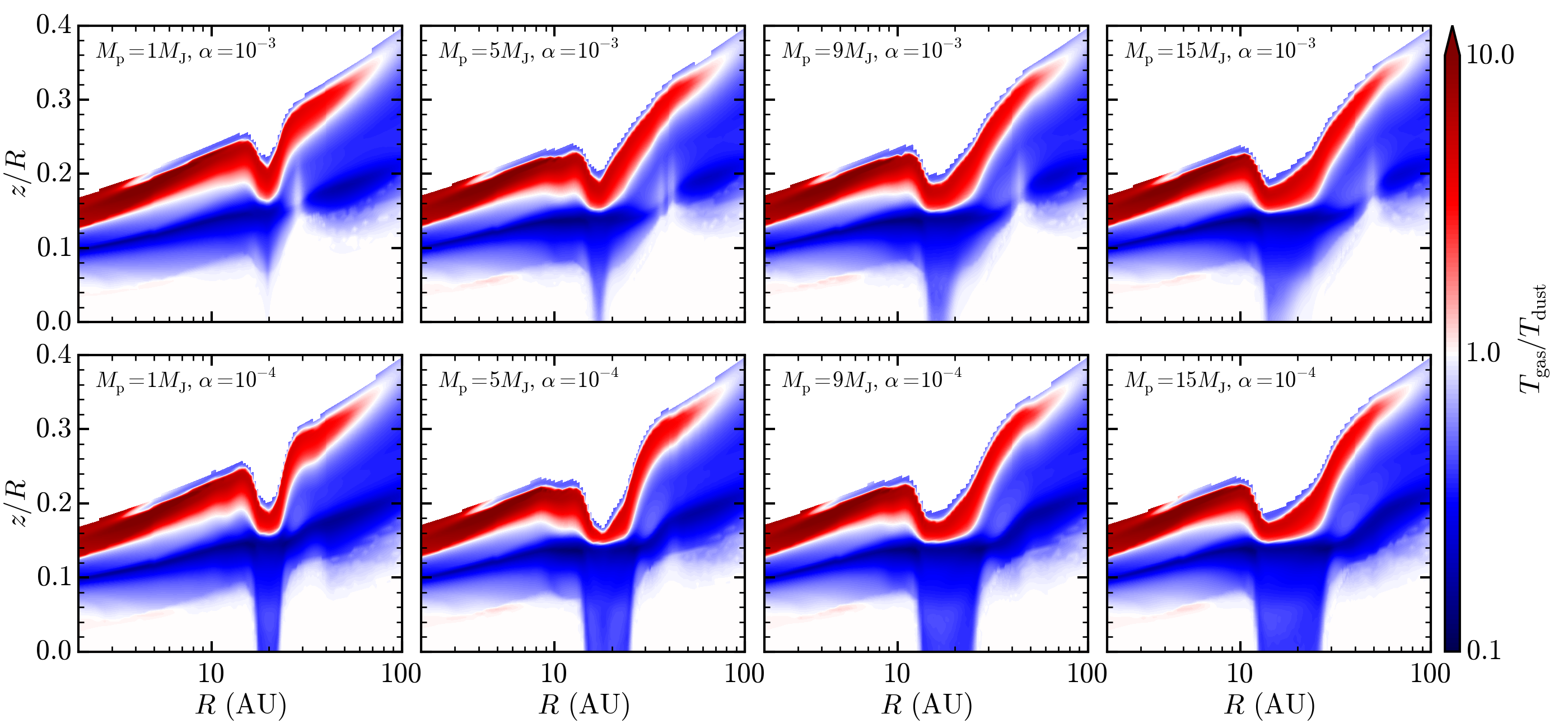}
\end{center}
\caption{Ratio of gas and dust temperatures for all models. Top and bottom panels represent $\alpha=10^{-3}$ and $10^{-4}$, respectively. From left to right, $M_{\rm p}=[1,\ 5,\ 9,\ 15]M_{\rm J}$.}
\label{fig:temp_ratio}
\end{figure*}

Finally, the gas temperature and chemical abundances are calculated with DALI computing the gas thermal balance iterated with time dependent chemistry for $1\,$Myr evolution. As in \citet{2017A&A...605A..16F} (see their Section 2.6 for more details), we accounted for the total dust surface area at every point in the disk explicitly in the thermochemical mechanisms depending on it, in particular the energy transfer in gas grains collisions, H$_2$ formation rate,  and freeze-out, desorption, and hydrogenation rates on the grain surfaces. Throughout this paper, a binding energy for CO of $855\,$K is assumed, as measured in laboratory experiments for pure CO ice \citep{1993ApJ...417..815S,2003ApJ...583.1058C,2005ApJ...621L..33O,2006A&A...449.1297B}. The abundance of PAHs (polycyclic aromatic hydrocarbons) is assumed to be $0.1$ of the typical ISM abundance in the whole disk \citep[][]{2012A&A...541A..91B}. Initial ISM-like elemental abundances are considered, where $[{\rm C}]/[{\rm H}]=1.35\times10^{-4}$, and $[{\rm O}]/[{\rm H}]=2.88\times10^{-4}$, and the notation $[X]$ indicates element $X$ in all its volatile forms. CO isotope selective photodissociation \citep[e.g.,][]{1988ApJ...334..771V} is not considered in this work. A relative abundance of $^{12}$CO/$^{13}$CO=70 and $^{12}$CO/C$^{18}$O=560 is assumed in the models, determined from the isotope ratios of $^{12}$C/$^{13}$C and $^{16}$O/$^{18}$O in the ISM \citep{1994ARA&A..32..191W}.

\begin{figure*}
\center
\includegraphics[width=.99\columnwidth]{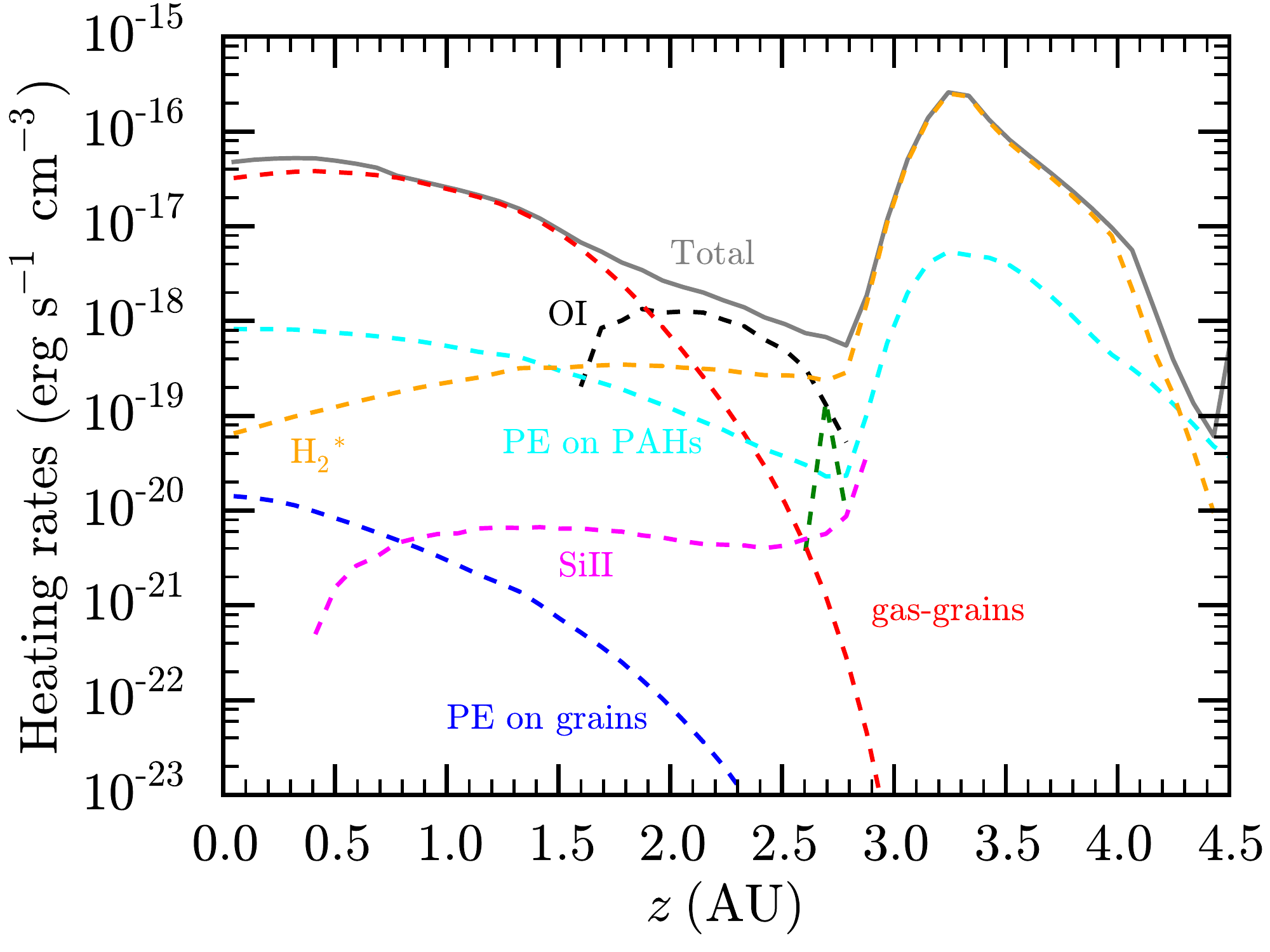}
\includegraphics[width=.99\columnwidth]{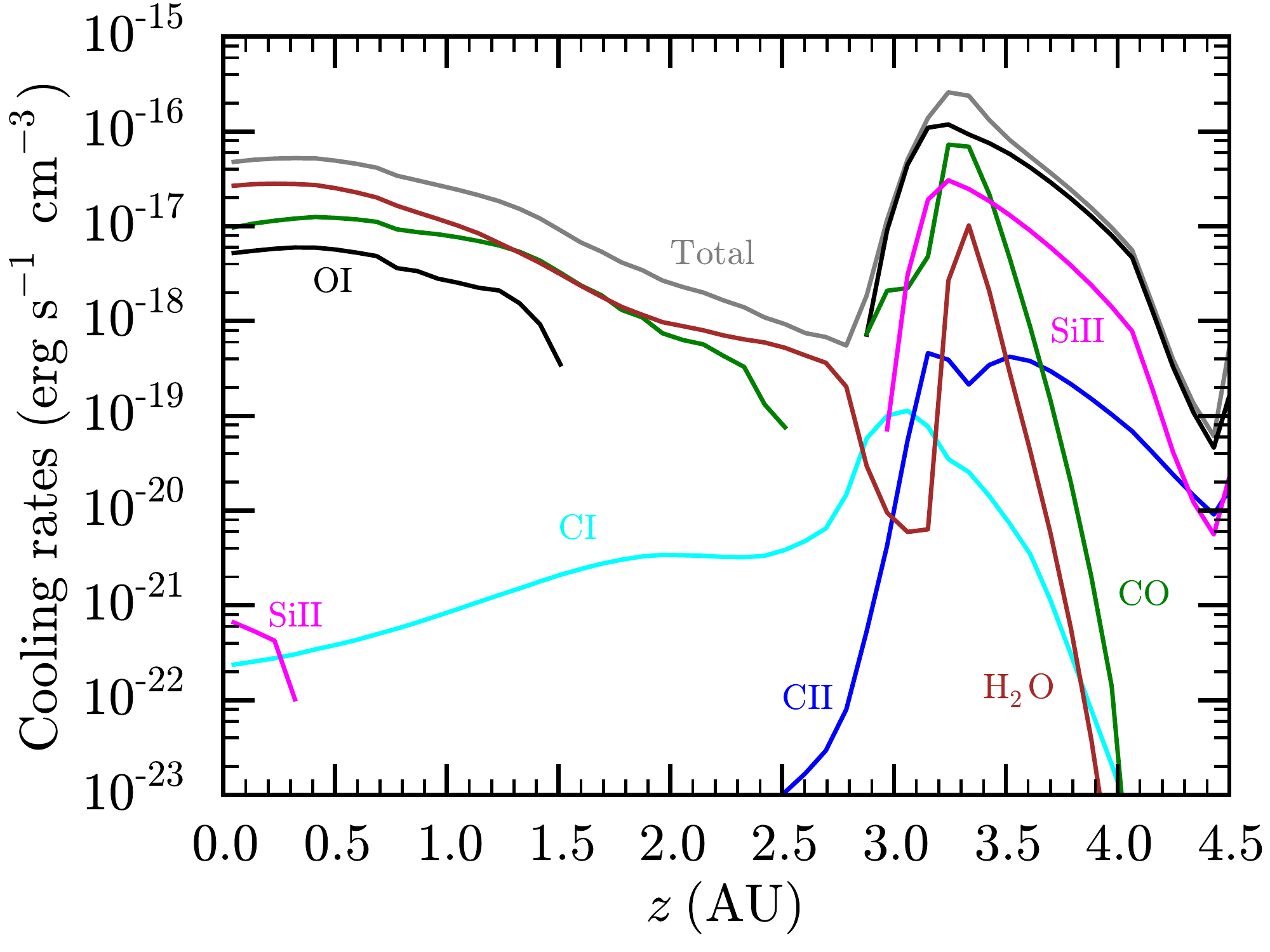}
\caption{Vertical cut of the main heating and cooling rates within the gap, at $20\,$AU, for the $\alpha=10^{-3}$, $M_{\rm p}=15\,M_{\rm J}$ model. Heating rates are shown with dashed lines; cooling rates are indicated with solid lines. Molecules and atoms contributing to both heating and cooling share the same color in both panels. The gray solid line shows the total cooling rate, which equals the total heating rate. Different colors indicate different contributors. In the legend, ``PE'' stands for photoelectric effect.
}
\label{fig:hc_functions}
\end{figure*}

In these calculations, we assume a stellar mass of $1\,M_\odot$, which is consistent with the hydrodynamics and dust evolution models. The stellar spectrum is a blackbody with a $4730\,$K surface temperature and a bolometric luminosity of $3\,L_\odot$. Ultraviolet excess due to accretion and possibly a hot component of the stellar photosphere is routinely observed toward T Tauri stars \citep[see review by][and references therein]{2016ARA&A..54..135H}. We consider an accretion rate $\dot{M}=10^{-9}\,M_\odot\,$yr$^{-1}$, where the gravitational energy is converted into photons distributed as a blackbody spectrum of $10000\,$K \citep{2016A&A...592A..83K}. The stellar radius is set to $1.7\,R_\odot$. This corresponds to an accretion luminosity $L_{\rm acc} = 1.8\times10^{-2}\,L_\odot$, and a UV excess of $L_{6-13.6\ {\rm eV}} = 1.4\times10^{-3}\,L_\odot$. The number of photon packages used in the radiative transfer is $3\times10^7$ to compute the dust temperature, and $3\times10^6$ in every wavelength bin to compute mean intensities \citep[details on the radiative transfer algorithm can be found in][]{2012A&A...541A..91B,2013A&A...559A..46B}. An external UV field of $1\,G_0$ is considered, where $G_0 \sim 2.7\times10^{-3}$\,erg\,s$^{-1}$\,cm$^{-2}$ is the UV interstellar radiation field between $911\,\AA$ and $2067\,\AA$ \citep{1978ApJS...36..595D}.

Finally, the (sub-)mm continuum and line emission images are calculated using the DALI ray tracer \citep{2012A&A...541A..91B}, assuming a distance of $150\,$pc, and a face-on disk.

\section{Results}

\label{sec:results}

\subsection{Density and temperature structure}
\label{sec:dens_temp}

The density structures recovered from the hydrodynamical and dust evolution simulations are shown in Fig.~\ref{fig:d2g_ratio} in the form of the local dust-to-gas ratio $\delta_{\rm dg}$ (an example of the actual gas and dust density distribution is shown in Appendix~\ref{app:abu}). As shown in \citet{2016MNRAS.459L..85D}, the planet carves a gap in the gas surface density in all simulations. In all cases, the gap opening criteria are satisfied \citep[e.g.,][where the latter is a recent review]{1993prpl.conf..749L,2006Icar..181..587C,2014prpl.conf..667B}. Given the dimensional parameters chosen in this paper, the so-called viscous criterion requires that $M_{\rm p}\gtrsim0.06M_{\rm J}$ (for $\alpha=10^{-3}$), whereas the thermal criterion requires that $M_{\rm p}\gtrsim M_{\rm th}=0.4M_{\rm J}$. Here the thermal mass $M_{\rm th}$ is defined as the mass at which the Hills radius of the planet is larger than the vertical scale height of the disk $H$. Given that the minimum mass considered is $M_{\rm p}=1M_{\rm J}$, a gap in the gas surface density is observed in all the simulations.

The planet torques induce a pressure maximum outside the planet location \citep[e.g.,][]{2004A&A...425L...9P}. \citet{2014A&A...572A..35L,2016MNRAS.459.2790R} have both found that a pressure maximum is generated whenever $M_{\rm p}\gtrsim0.2M_{\rm th}$, which is indeed the case in the simulations presented here. For lower mass planets that are not able to create a pressure maximum, the simultaneous evolution of the dust and gas hydrodynamical variables should be tracked to compute the dust surface density distribution \citep[e.g.,][]{2015MNRAS.453L..73D,2016MNRAS.459L...1D,2016MNRAS.459.2790R}, which we do not consider in this work. The steepness and amplitude of the pressure gradient regulates the radial migration of the dust particles, depending on their sizes \citep[e.g.,][]{2012A&A...538A.114P}. The net effect is that particles with sizes $\gtrsim100\,\mu$m in the outer regions of the disk tend to accumulate at the location of the pressure maximum, yielding a so-called dust trap. In the $\alpha=10^{-3}$ cases the dust trap is clearly visible for all planet masses, where the radial location of the dust trap increases with mass of the perturbing planet \citep[e.g.,][]{2013A&A...560A.111D}. The dust trap is also present in the $\alpha=10^{-4}$ case, but is less apparent in Fig.~\ref{fig:d2g_ratio} since the low turbulence leads to a growth to particle sizes larger than $1\,$cm, which are heavily settled toward the disk midplane and have very low opacities. Details of the radial distribution of the grain size distribution can be found in fig.~1 of \citet{2016MNRAS.459L..85D}. In the low turbulence ($\alpha=10^{-4}$) cases, the gap in the gas surface density generated by the planet torques is almost devoid of dust. The width of this gap as shown in the dust-to-gas ratio $\delta_{\rm dg}$ depends on the mass of the planet, as expected from many numerical simulations \citep[e.g.,][]{1986ApJ...309..846L,1999ApJ...514..344B,2006Icar..181..587C}. The dust in the low viscosity cases is always more vertically settled than in the $\alpha=10^{-3}$ cases because of the lower turbulent velocities that are inefficient in stirring up the dust grains and the more effective growth.

\begin{figure}
\center
\includegraphics[width=\columnwidth]{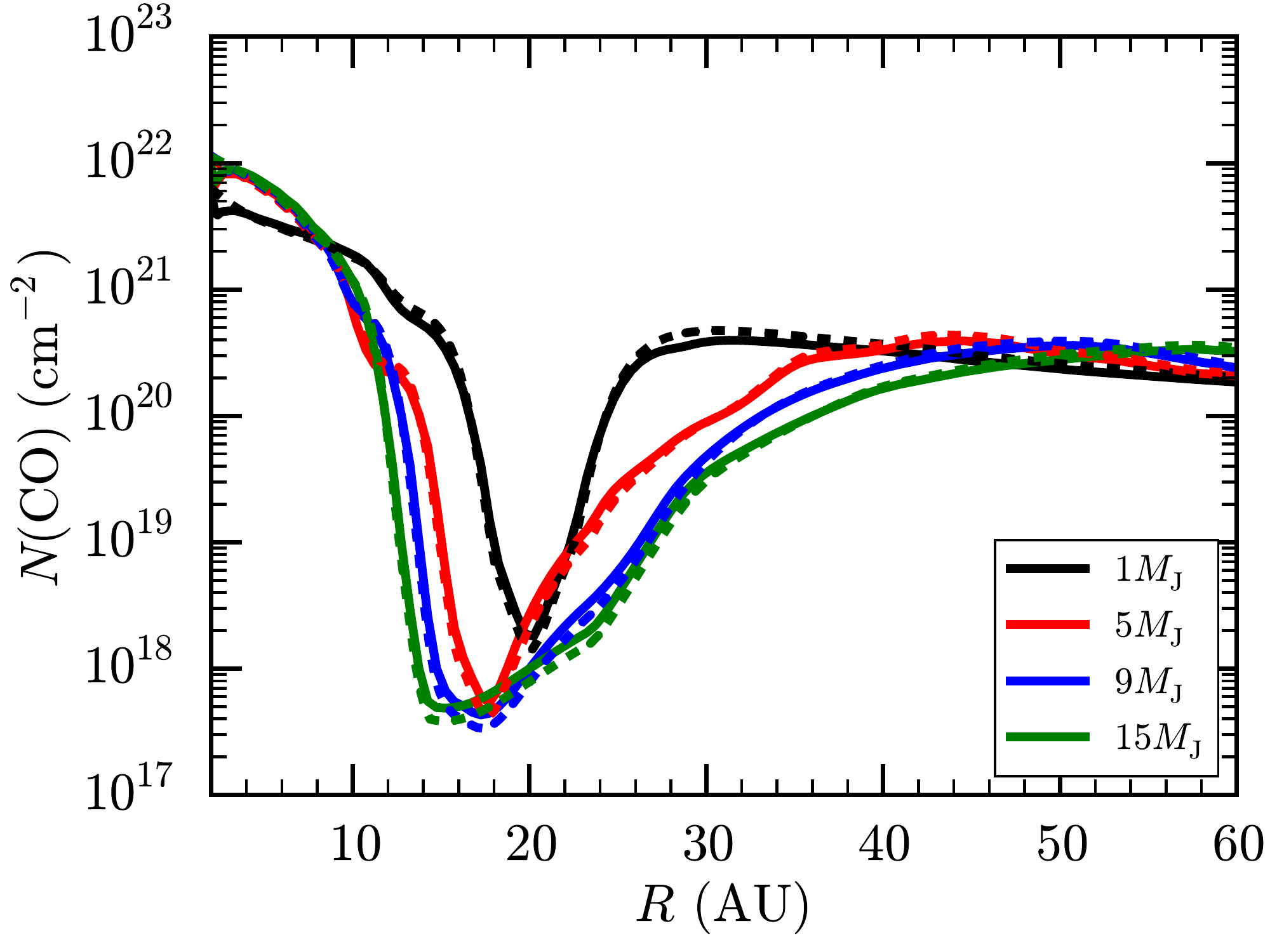}
\caption{Solid lines: CO column densities for  $\alpha=10^{-3}$ cases. The dashed lines indicate total gas surface densities normalized to the CO column densities at 5\,AU, where the normalization factor is on the order of the total carbon abundance assumed in the models.
}
\label{fig:co_col}
\end{figure}

The low dust surface density in the proximity of the planet orbit increases the physical penetration depth of UV photons within the gap opened by the planet in the gas surface density (see Appendix \ref{app:temp} for more details). As for the gas and dust density structure, the radial width of the region in which UV photons can penetrate almost to the disk midplane increases with planet mass and decreases with disk turbulence. The deeper penetration of the stellar radiation increases the dust temperature by a few degrees, when compared to a smooth surface density profile (see Appendix \ref{app:temp} for detailed figures). We note however that the UV flux within the gap is $\lesssim1\,G_0$, thus not high enough to provide substantial heating to the gas via photoelectric effect on PAHs and small grains. The low dust-to-gas ratio in the gaseous gap opened by the planet introduces a thermal decoupling between the gas and dust component. As shown by \citet{2017A&A...605A..16F}, the energy transfer between gas and dust grains depends strongly on the total dust surface area available and in regions where this is indeed low the gas component can become colder than the dust. This is clearly shown in Fig.~\ref{fig:temp_ratio}, where the ratio of the gas and dust temperatures becomes lower than unity in the proximity of the planet location. This effect is more prominent in the low viscosity case, and in the most massive planet simulations, where the planet torques can carve a well-defined gap in the dust surface density. This is different from the study by \citet{2013A&A...559A..46B}, where the effects of dust growth were not included.

Since photoelectric heating is low, the most important heating function in the disk midplane is collisions between gas and dust particles (see Fig.~\ref{fig:hc_functions}). In the upper layers of the disk around the planet location at $20\,$AU, vibrationally excited H$_2$ (by FUV pumping) and photoelectric heating of PAHs are the dominant heating functions \citep[as found by][for $R\gtrsim20\,$AU in their models]{2013A&A...559A..46B}. As for the coolants, in the disk upper layers, atomic oxygen, CO, and Si\,\textsc{ii} are the most important cooling species, whereas both atomic carbon and C\,\textsc{ii} do not contribute significantly. In the disk midplane, the main coolants in the models are CO, H$_2$O, and O\,\textsc{i}. The abundances of volatile H$_2$O and atomic oxygen are found to be high in the gap, since most of the oxygen is not sequestered in water ice as a result of the very low dust surface area available to freeze on (see Appendix \ref{app:abu} for more details).

\begin{figure*}
\begin{center}
\includegraphics[width=.85\textwidth]{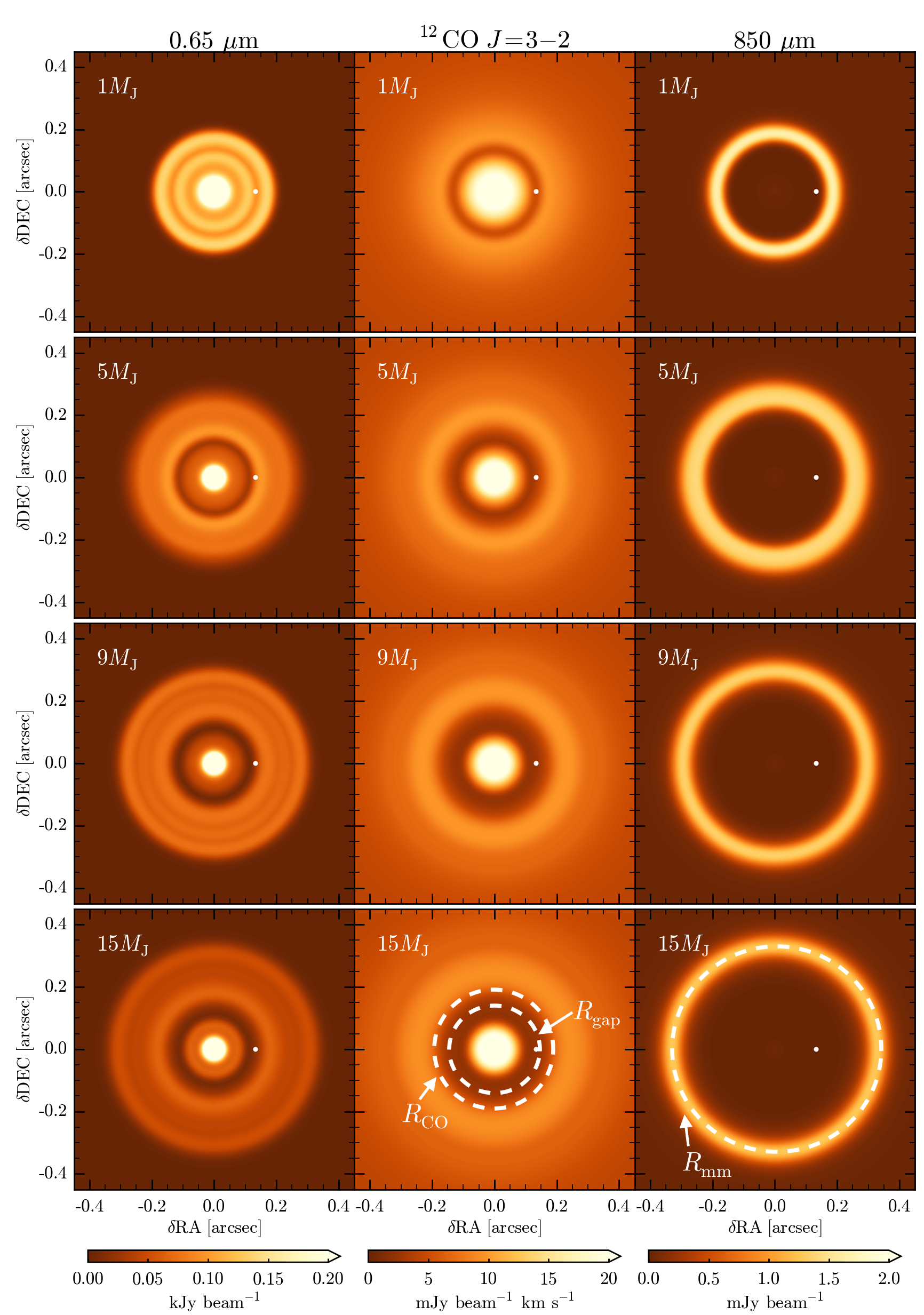}
\end{center}
\caption{R-band ($0.65\,\mu$m) polarimetric emission map, moment 0 map of the  $^{12}$CO $J$=3-2 line, and continuum emission map at $850\,\mu$m for the $\alpha=10^{-3}$ simulations. The resolution is set to $0.03\arcsec$; the disk is located at a distance of $150\,$pc. The white dot in each panel indicates the distance of the planet from the central star. The bottom row shows the quantities $R_{\rm gap}$, $R_{\rm CO}$, and $R_{\rm mm}$, which are defined in Section~\ref{sec:mass_p}. 
}
\label{fig:maps_alpha3}
\end{figure*}

\begin{figure*}
\begin{center}
\includegraphics[width=0.49\textwidth]{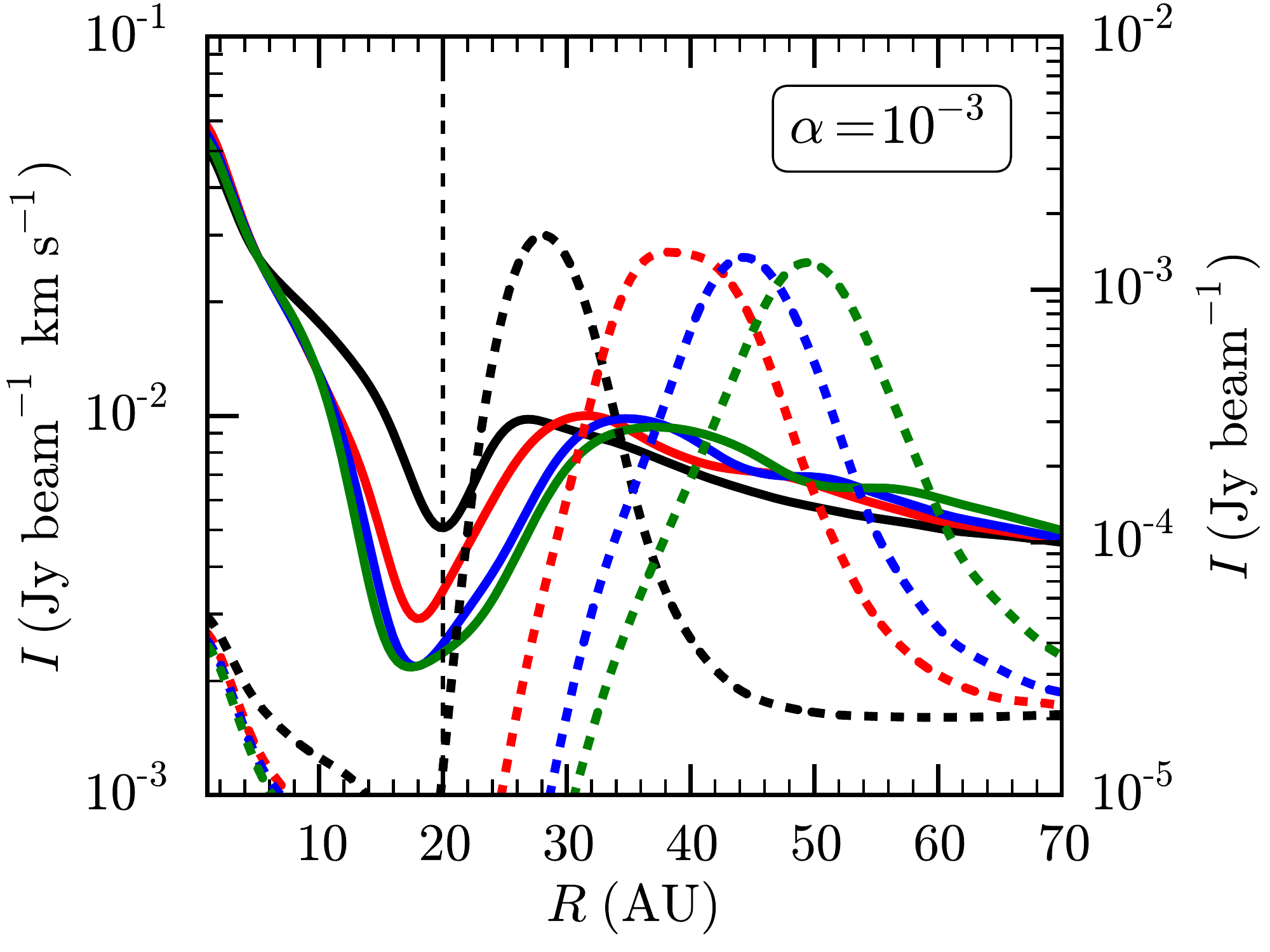}
\includegraphics[width=0.49\textwidth]{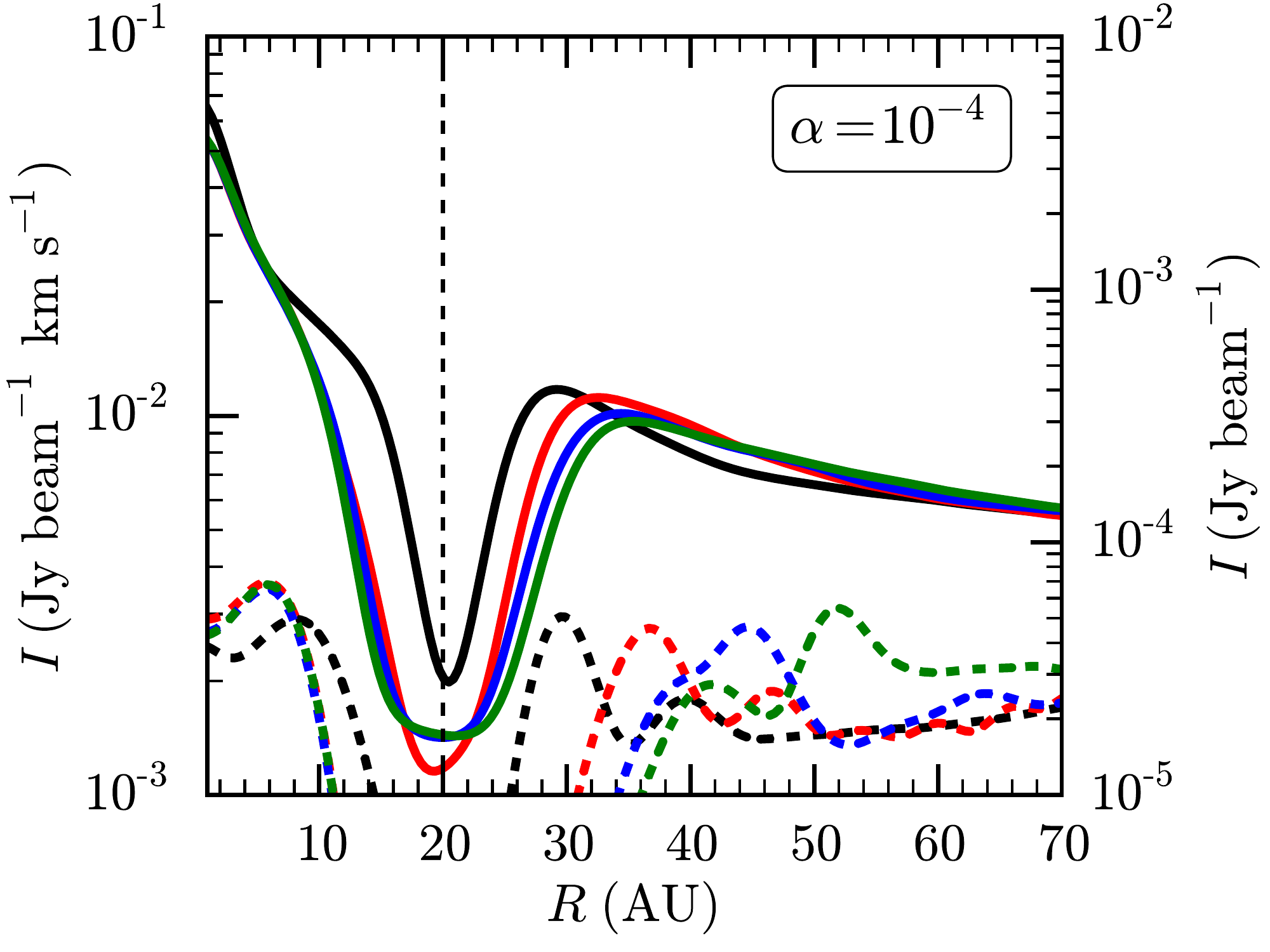}
\end{center}
\caption{Radial intensity profiles of the $^{12}$CO $J$=3-2 line moment 0 map (solid lines) and of continuum at $850\,\mu$m (dashed lines) of all models, as in Fig.~\ref{fig:maps_alpha3}. The $y$-axis for the continuum emission is on the right axis of the plots; color coding is as in Fig.~\ref{fig:co_col}. The full continuum profiles can be found in \citet{2016MNRAS.459L..85D}.
}
\label{fig:profiles_from_maps}
\end{figure*}

\begin{figure*}
\begin{center}
\includegraphics[width=0.32\textwidth]{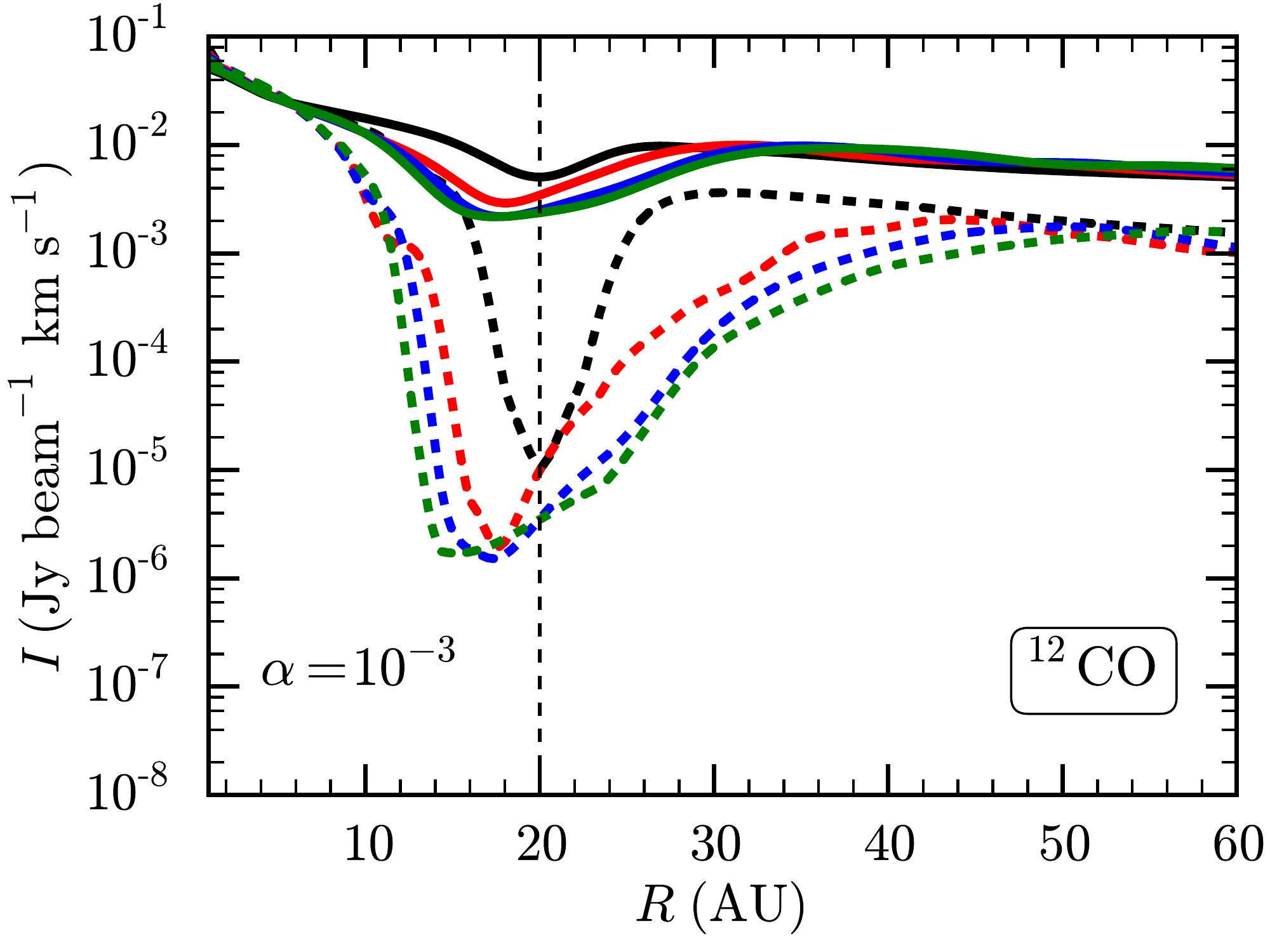}
\includegraphics[width=0.32\textwidth]{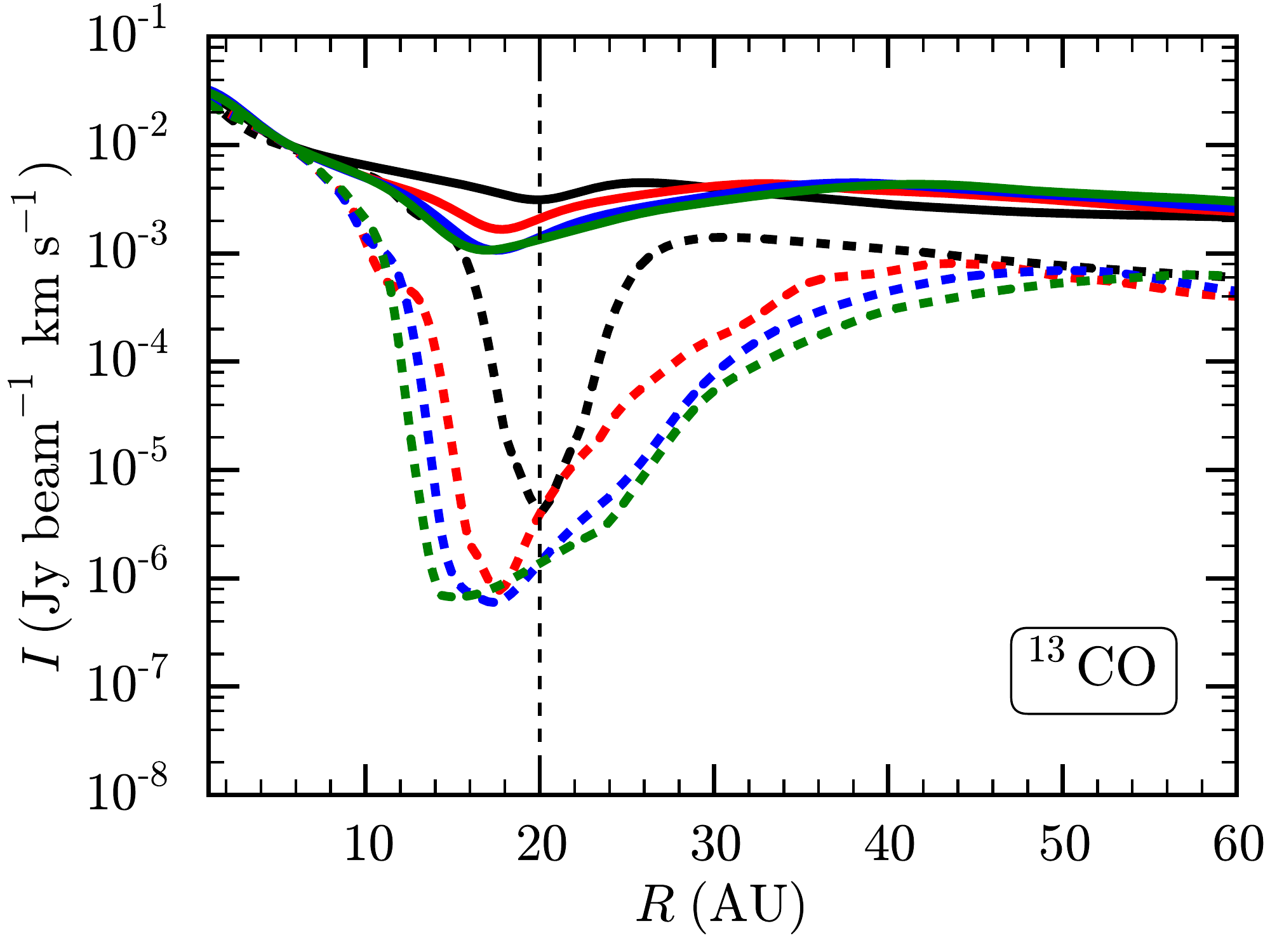}
\includegraphics[width=0.32\textwidth]{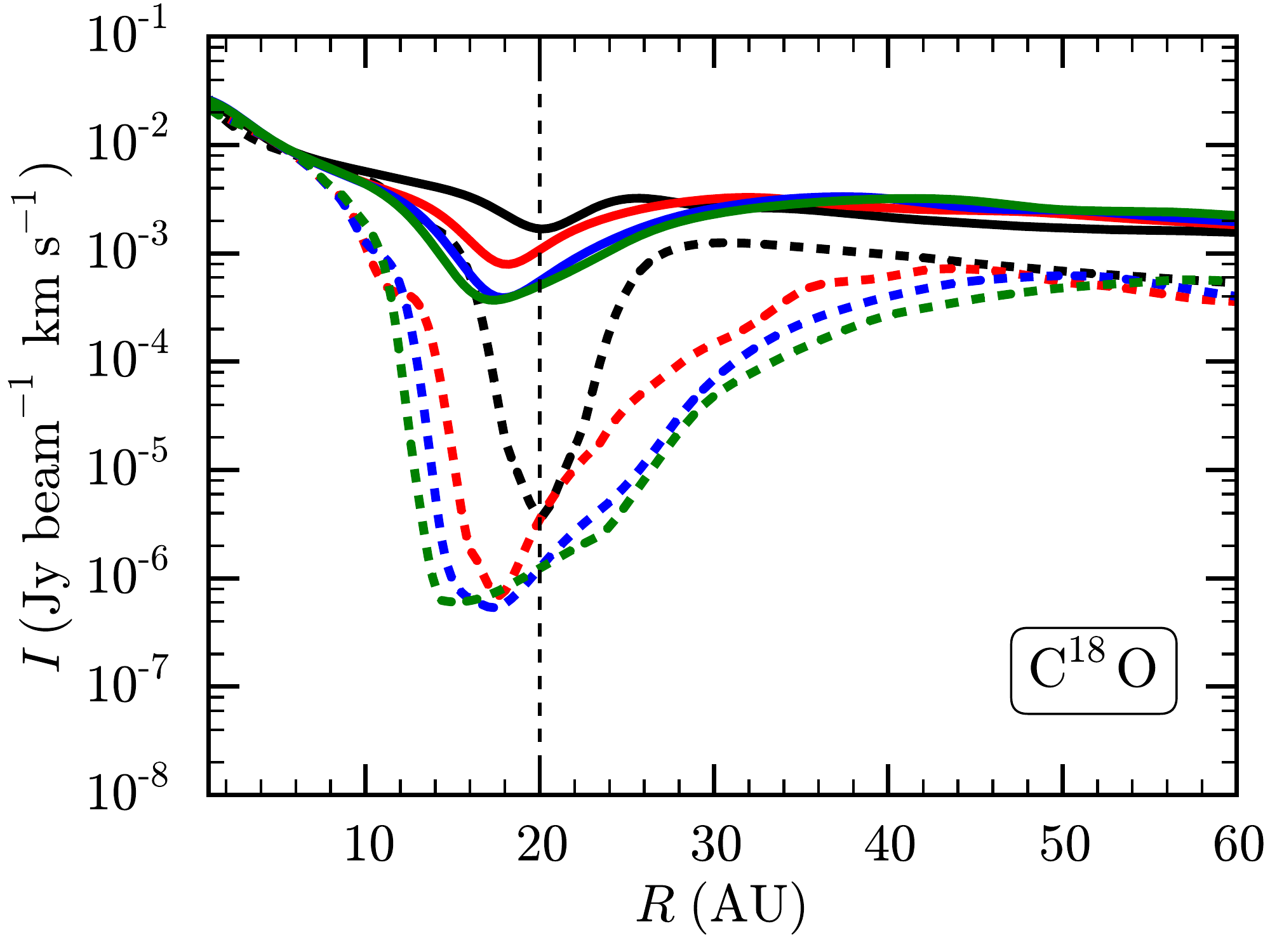}\\
\includegraphics[width=0.32\textwidth]{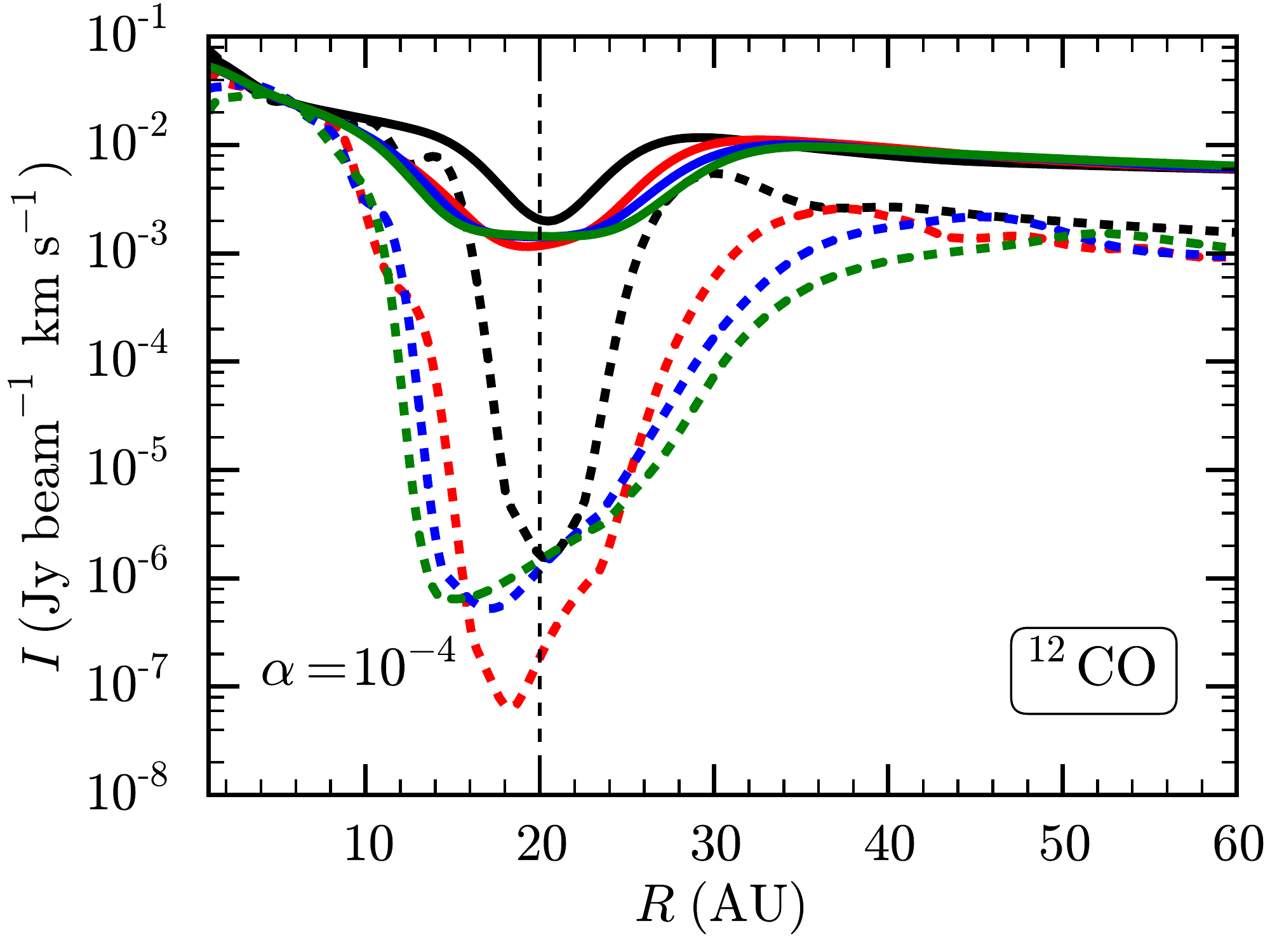}
\includegraphics[width=0.32\textwidth]{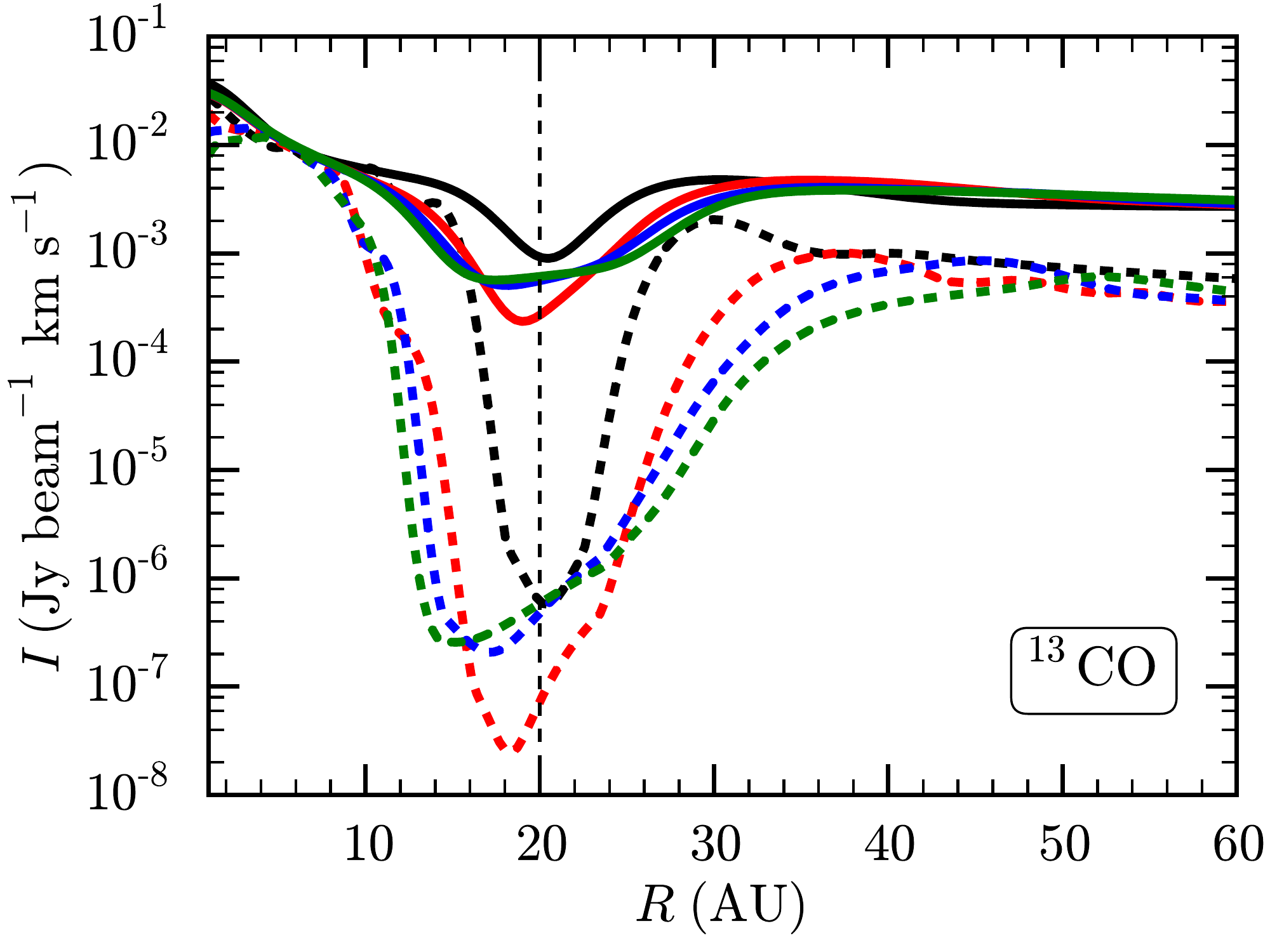}
\includegraphics[width=0.32\textwidth]{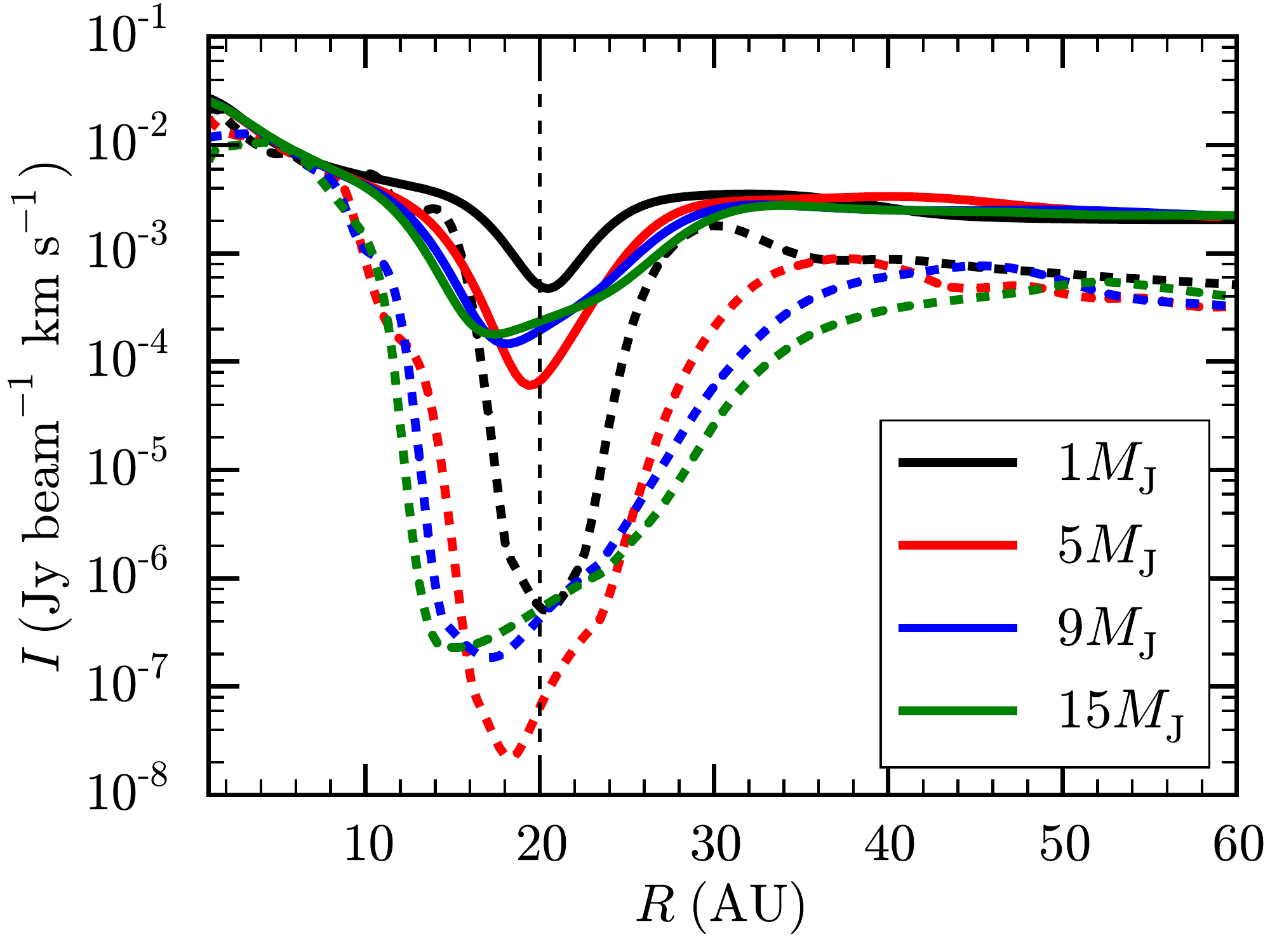}\\
\end{center}
\caption{Radial intensity profiles of $J$=3-2 CO isotopolog emission lines (solid lines). In the top and bottom panels, $\alpha=10^{-3}$ and $10^{-4}$, respectively. From left to right: $^{12}$CO, $^{13}$CO, C$^{18}$O. The emission was convolved with a $0.03\arcsec$ resolution beam for disks located at a distance of $150\,$pc. The dashed lines show the gas surface densities from the hydrodynamical simulations normalized at the emission at $\sim5\,$AU. The vertical dashed line indicates the radial location of the planet.
}
\label{fig:profile_surf_dens_01}
\end{figure*}

The vertical column density of CO in the gas phase resembles very well the total gas surface density, even in the simulations with very massive planets (see Fig.~\ref{fig:co_col}). This indicates that even though the UV field can penetrate in the gas gap, the amount of photodissociated CO within the gap is not significant (less than $1\%$). The reason is CO self-shielding, which operates effectively for CO column density $\gtrsim10^{15}\,$cm$^{-2}$ \citep{1988ApJ...334..771V}. We note that \citet{2013A&A...559A..46B} found that CO can survive within transition disk cavities even when directly illuminated by more intense UV radiation thanks to the same mechanism. The emitting surface layer of all CO isotopologs is in LTE, given the very low critical density of rotational transitions of CO. Even within the gaps subthermal excitation is never observed in these models.

\subsection{Continuum versus CO emission}

\begin{figure*}
\begin{center}
\includegraphics[width=0.45\textwidth]{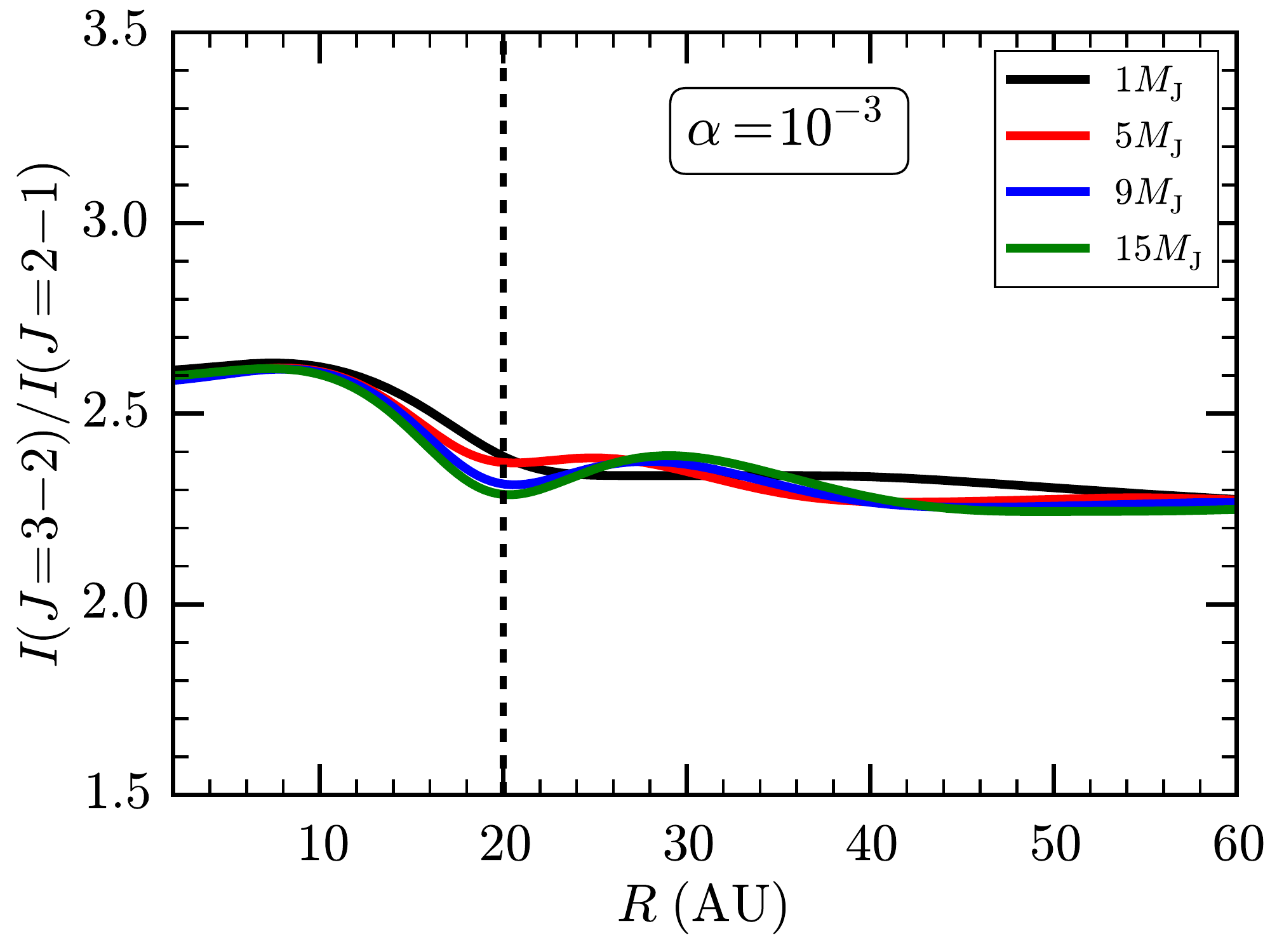}
\includegraphics[width=0.45\textwidth]{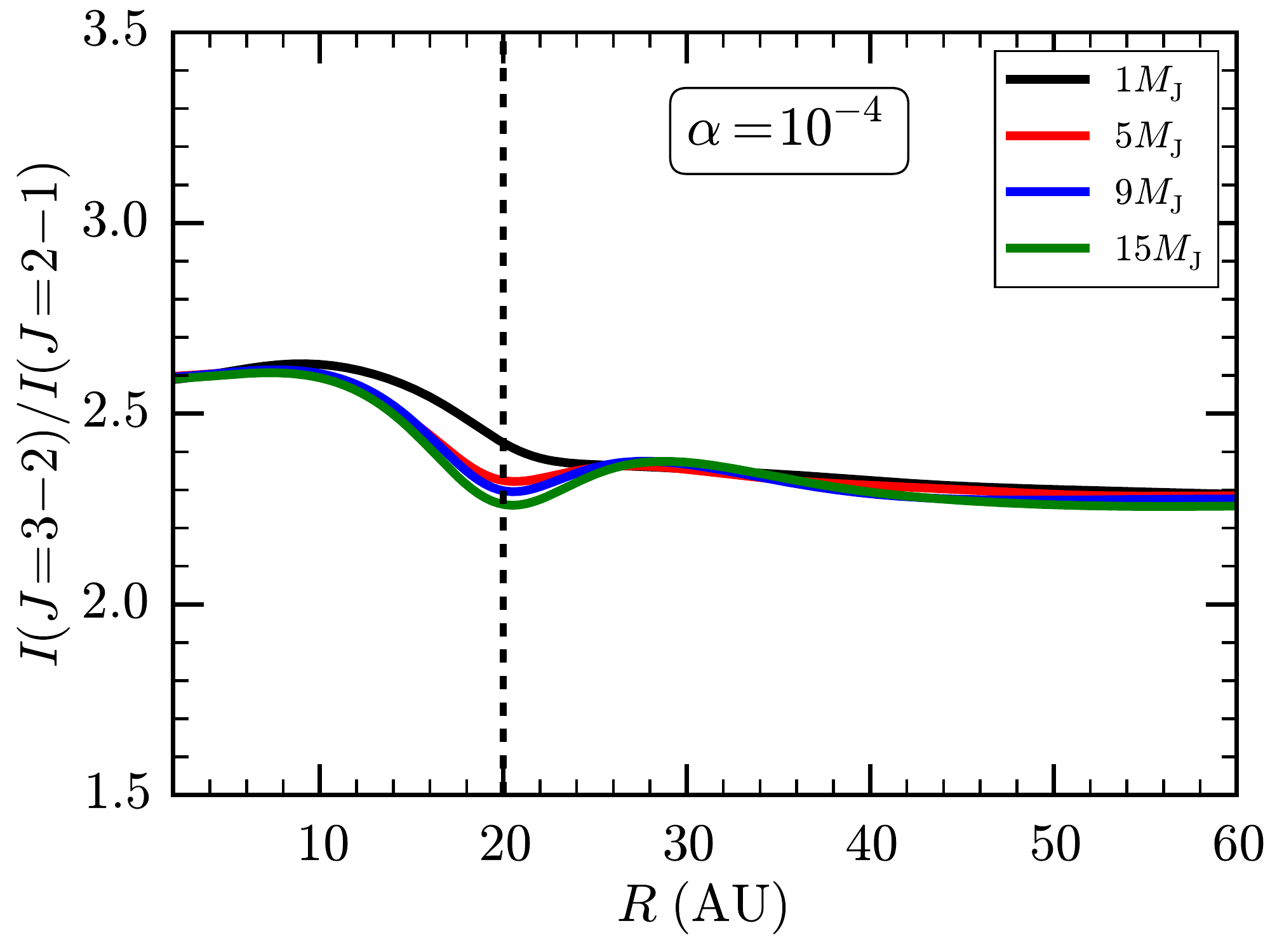}\\
\includegraphics[width=0.45\textwidth]{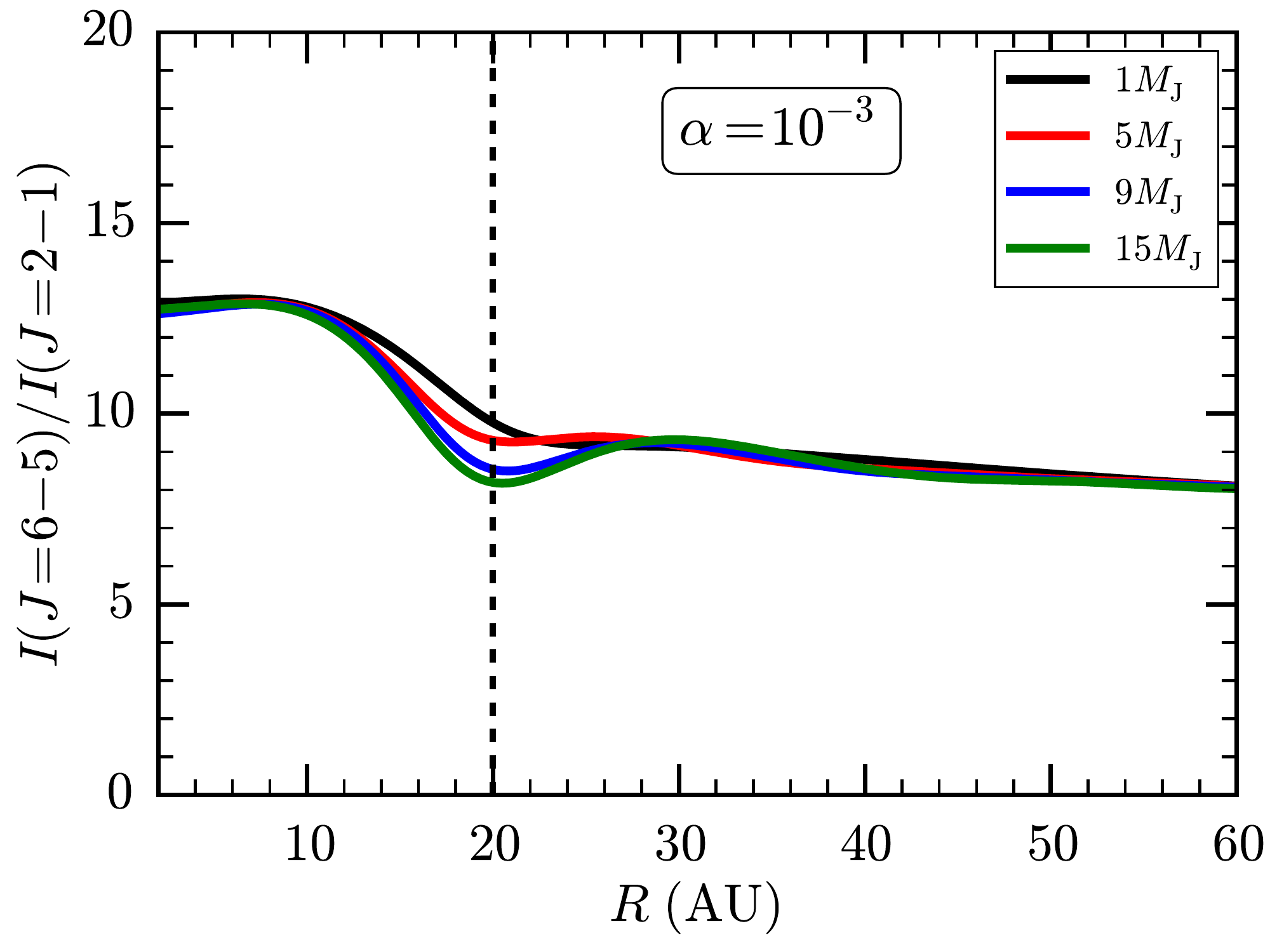}
\includegraphics[width=0.45\textwidth]{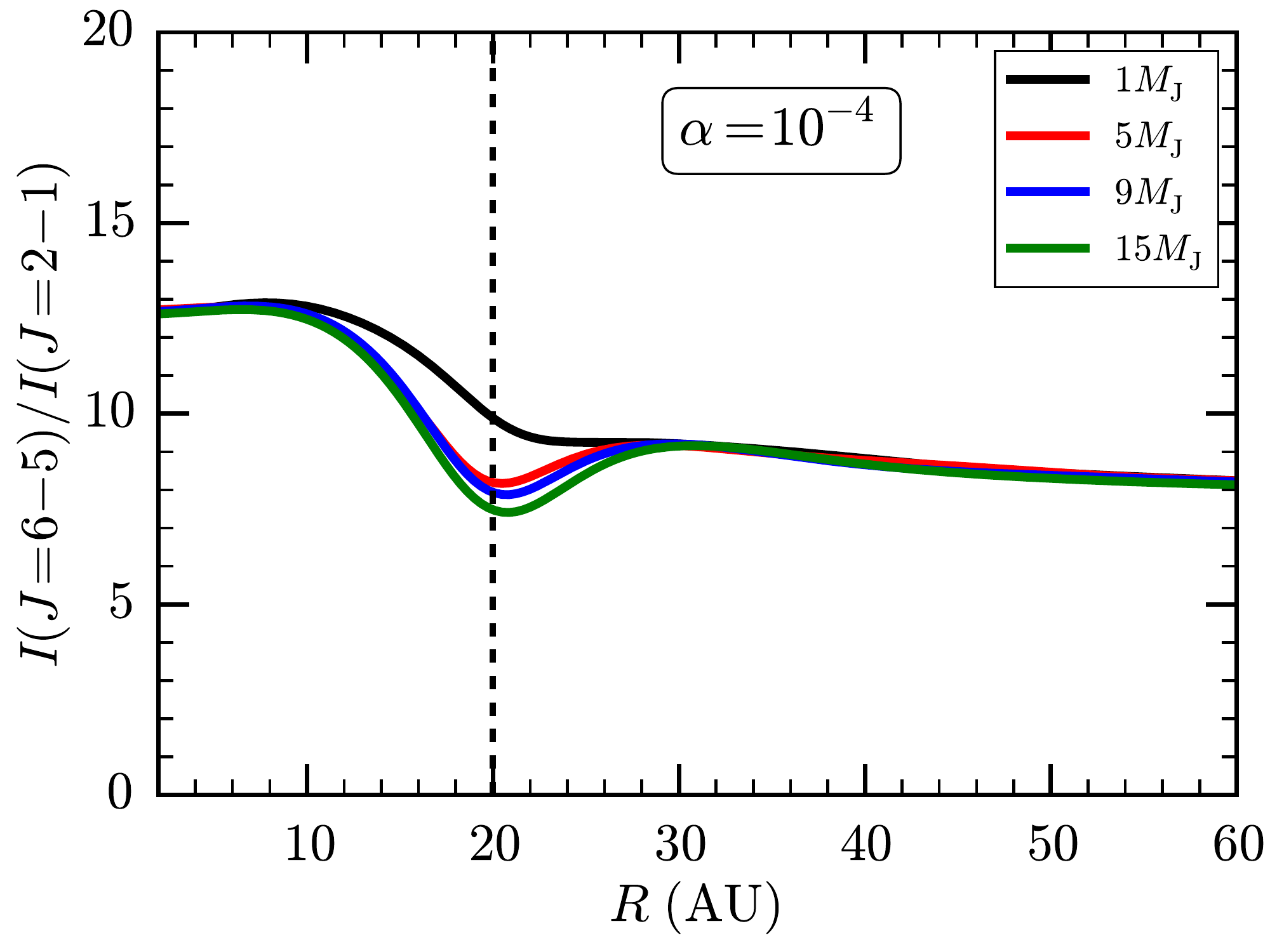}
\end{center}
\caption{Line ratios for $^{12}$CO. The emission was convolved with a $0.07\arcsec$ resolution beam for sources located at a distance of $150\,$pc. The vertical black dashed line indicates the radial location of the planet. The CO $J$=6-5/2-1 line ratios shows the effects of the varying gas temperature much more clearly than the CO $J$=3-2/2-1 line ratios.
}
\label{fig:line_ratios}
\end{figure*}

The dust and gas density distributions, together with the radiation field and temperature structures, concur in determining the emission in the (sub-)mm continuum and low rotational lines of CO isotopologs, in particular $^{12}$CO, $^{13}$CO and C$^{18}$O. Fig.~\ref{fig:maps_alpha3} summarizes the intensity map of the $^{12}$CO $J$=3-2 line and continuum at $850\,\mu$m of all models with $\alpha=10^{-3}$, as obtained from the DALI ray tracer. The same figure for the $\alpha=10^{-4}$ models is in the Appendix (see Fig.~\ref{fig:maps_alpha4}). In the same figures, we also show the R-band polarimetric emission maps, which are taken from \citet{2016MNRAS.459L..85D},  to compare CO, scattered light, and sub-mm continuum emission. Together these maps provide complementary information on the gas density and temperatures the small versus large grain distribution. The R-band polarimetric emission maps  have been generated with MCMax \citep{2009A&A...497..155M}, since DALI does not include polarization in the continuum radiative transfer. In this case, the azimuthally averaged dust and gas radial profiles are imported into MCMax, where the gas vertical structure of the disk is solved iteratively assuming hydrostatic equilibrium. Dust vertical settling is also computed directly by MCMax, where turbulent mixing with the $\alpha$ viscosity used in hydro simulations is considered. More details on the method can be found in section 2.2 of \citet{2013A&A...560A.111D}.

In the $\alpha=10^{-3}$ simulations, mm-sized particles are effectively accumulated in the dust trap, and thus the emission shows clear rings outside of the planet location. The position of such ring traces the position of the pressure maximum, which moves outward with planet mass. In the $\alpha=10^{-4}$ simulations, the general trend is similar to the higher turbulence cases. However, the low turbulence has led the grain size distribution to be significantly dominated by large grains ($\gtrsim1\,$cm). The opacities of these large particles even at mm wavelengths is rather low and the flux at $850\,\mu$m is lower than the analog simulations with $\alpha=10^{-3}$ by a factor of $\sim10$. The contrast between the outer ring and the inner disk emission at $850\,\mu$m is thus significantly reduced and the inner disk shows the same level of surface brightness as the outer ring (see Fig.~\ref{fig:profiles_from_maps}). We note that the wiggles observed in the mm emission of the $\alpha=10^{-4}$ simulations are an artifact due to the azimuthal average of the gas hydrodynamical simulations, where the spirals launched by the planet are artificially mapped into secondary local maxima in 1D \citep[see][for more details]{2016MNRAS.459L..85D}.

The $^{12}$CO $J$=3-2 emission maps show a clear gap at the location of the planet in all simulations. Both the depth and width of this gap increase with planet mass. The depth of the gap decreases with turbulence, whereas its width does not depend strongly on $\alpha$. Significantly, in all the models explored here, the CO emission always shows a gap, rather than a cavity; the surface density in the inner regions is very bright owing to the warmer temperatures in the inner disk. Thus the CO emission seems to well probe the gas surface density of the simulations, and the gas surface density from the hydrodynamical simulations increases as it approaches the central star (see Fig.~\ref{fig:co_col}). The $^{12}$CO $J$=2-1 emission maps look very similar to the $^{12}$CO $J$=3-2 maps, where the gap in CO is slightly less prominent (see discussion in Section \ref{sec:line_ratio}).

Finally, in all simulations the scattered light images resemble the CO emission maps, even though the scattered light images show a more complex structure. In general, the comparison indicates that the small dust may be used as a tracer of the underlying gas surface density when such a deep gap is present, even though we must bear in mind many caveats  when trying to extrapolate from scattered light observations to gas surface densities.
The comparison between R-band and (sub-)mm continuum maps is discussed in more detail in \citet{2016MNRAS.459L..85D}.

\subsection{CO isotopologs intensity profiles: Surface density and temperature gaps}
\label{sec:line_ratio}

Fig.~\ref{fig:profile_surf_dens_01} shows the radial intensity profiles in the $J$=3-2 rotational transition for the three main CO isotopologs ($^{12}$CO, $^{13}$CO, and C$^{18}$O) for all models, compared with the relative gas surface density profiles from the hydrodynamical simulations. The dynamic range of the gap in the CO emission (i.e., the depth of the gap) becomes larger with planet mass, saturating at $M_{\rm p}\gtrsim5\,M_{\rm J}$, as seen in the gas surface density (see also Fig.~\ref{fig:co_col}). Instead, the width of the CO gap increases with planet mass up to the very end of the masses explored here. The shape of the emission profile of all isotopologs traces the surface density profile, but the gap depth is much less deep in emission than in the underlying surface density. The dependence of the CO emission on the surface density is roughly logarithmic for all cases, indicating that the emission of all isotopologs is (partly) optically thick even within the gaps. 
For the most optically thick CO isotopolog, $^{12}$CO, the gap is $\sim1$ order of magnitude deep. For rarer isopologues, in particular C$^{18}$O, the depth of the gap as traced by line emission becomes larger ($\sim2$ orders of magnitude). For lower disk masses, or for lower surface densities where the planet opens the gap, the gap in the C$^{18}$O emission should be even deeper, as the emission line becomes optically thinner due to the lower column densities.

In Fig.~\ref{fig:temp_ratio} we show that the gas can become thermally decoupled from the  dust  within the gaps. Since in almost all cases the emission is still optically thick for both $^{12}$CO and $^{13}$CO, the gas temperature is important in determining the radial intensity profile of the CO rotational lines. This can easily be probed by looking at line ratios, shown in Fig.~\ref{fig:line_ratios}. The CO gaps can be identified in the $J$=3-2 to $J$=2-1 line ratio; the line ratio shows a decrease of $\sim30\%$ at the planet location. The effect is even more prominent in the $J$=6-5 to $J$=2-1 line ratio and there is a decrease of $\sim50\%$ at the planet location. These results show that the gaps in CO emission are due to a combination of a gap in both surface density and temperature. We note that the cooling timescale of the gas is shorter than the advection timescale of the material through the gaseous gap, thus the gas flowing through the gap carved by the planet has time to cool to the lower temperatures deduced from the thermochemical models. The ALMA simulated observations of all models of the $^{12}$CO $J$=3-2 and $J$=6-5 lines are shown in Appendix \ref{app:ALMA}.

As shown in Fig.~\ref{fig:profile_surf_dens_01}, the gap is present in all models in the highly optically thick rotational lines of $^{12}$CO. In contrast, \citet{2015A&A...579A.105O} showed that the gap is not seen in emission maps of optically thick molecular lines (see their fig.~3). This highlights how important the gas thermal structure is when predicting emission lines of optically thick molecules, where the brightness temperature varies linearly with the excitation temperature of the molecule, and has a logarithmic dependence on the molecular column density. \citet{2015A&A...579A.105O} assumed perfect density and temperature coupling between gas and dust and computed dust temperatures assuming a uniform grain size distribution across the gap. Since the dust (and thus gas) temperatures are not varying significantly across the gap in their models, the line surface brightness does not show a gap. Allowing a thermal decoupling between gas and dust, as done here, significantly increases the detectability of gaps in optically thick molecular lines.

\section{Discussion}
\label{sec:discussion}

\subsection{CO gap versus CO cavity: A resolution effect?}

\begin{figure}
\center
\includegraphics[width=\columnwidth]{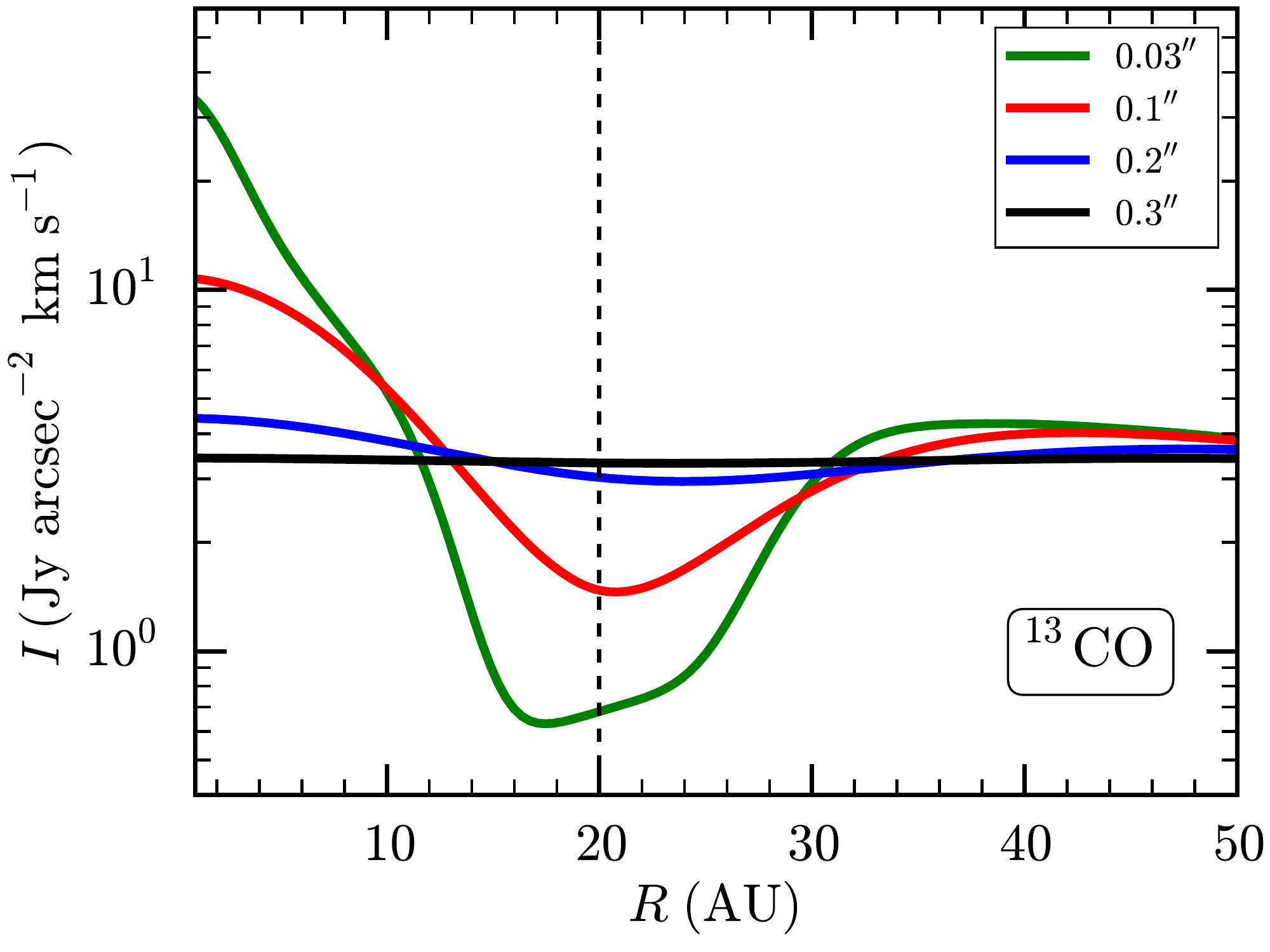}
\caption{Radial intensity profile of the $^{13}$CO $J$=3-2 line of the $\alpha=10^{-4}$, $M_{\rm p}=15\,M_{\rm J}$ simulation with different convolution beams. The source is located at $150\,$pc. The vertical black dashed line indicates the radial location of the planet. Even at the lower resolution, the CO emission does not show a cavity, but clearly shows a gap.}
\label{fig:res}
\end{figure}

In Section \ref{sec:results} we showed how at intermediate - high angular resolutions ($0.1\arcsec-0.03\arcsec$) for a source located at $150\,$pc the CO emission maps clearly show a gap located at the planet location. However, ALMA observations of the brightest transition disks have revealed that many of them show a cavity \citep{2015A&A...579A.106V,2016A&A...585A..58V,2017ApJ...836..201D}, as probed in particular by rotational lines of $^{13}$CO, which is optically thinner than $^{12}$CO, but still bright enough to provide good signal-to-noise observations. One possibility would be that the available observations do not have high enough angular resolution to see a gap, and the inferred cavity is just the dip in emission in the gap spread over the inner region of the disks. This possible scenario is tested in Fig.~\ref{fig:res}, in which the intensity profile of the $^{13}$CO $J$=3-2 line is shown for different convolution beams. The simulation chosen for this analysis is the low viscosity, $M_{\rm p}=15M_{\rm J}$ case, where the dip in CO emission is the deepest. Even for an angular resolution of $0.3\arcsec$, the molecular intensity profile never shows an actual cavity, indicating that some additional mechanism not included needs to reduce the CO emission from the very inner regions to reproduce the observations. Moreover, as discussed in Section~\ref{sec:intro}, the low CO column densities in the inner disks of some transition disks are also probed by the modeling of CO vibrationally excited IR lines \citep{2008ApJ...684.1323P,2014A&A...567A..51C,2017A&A...598A.118C,2015A&A...574A..75V,2015ApJ...809..167B,2017ApJ...836..242D}; this thus corroborates the evidence of depletion of CO in the inner disk regions, which is not compatible with the gap scenario. If accretion onto the planet were included in the hydrodynamical simulations, the gas surface density in the inner regions would have been lower by a factor of a few at the most \citep[e.g.,][see their fig.~9]{2016PASA...33....5O}, and the ratio of accretion onto the planet and onto the star is of order unity in most simulations \citep[e.g.,][]{2006ApJ...641..526L}. The slightly lower surface density would not reduce the $^{12}$CO and $^{13}$CO emission in the low $J$ lines significantly from the very inner regions of our models, since  relatively close to the central star such lines are highly optically thick.

\subsection{Using simultaneous ALMA gas and dust observations to probe planet masses}
\label{sec:mass_p}

Semi-analytical relations linking the properties of dust gaps as seen in scattered light of (sub-)mm continuum with planet masses have been presented in the literature (see Section \ref{sec:intro} for references), mostly relying on the combination of optical/IR data with (sub-)mm data. However, similar relations can been obtained comparing the (sub-)mm continuum emission with the CO intensity maps. We note that by using CO as a gas tracer, we do not need to rely on the scattered light observations to trace the gas surface density. Moreover, low $J$ CO emission lines and the (sub-)mm continuum can be obtained simultaneously with ALMA observations, thereby simplifying the analysis procedure. Finally, gas tracers can probe regions much deeper into the disk than scattered light images.
In this work, the gap depth is not advocated as a probe. This observational diagnostic is not easily derived from CO measurements, in particular because of optical thickness and excitation effects (since the gap might be a temperature gap as well, see Fig.~\ref{fig:temp_ratio}), which complicate the picture when trying to determine the gas depth from bright isotopologs. We emphasize that when estimates of the depth of observed CO gaps \citep[e.g.,][]{2016A&A...585A..58V,isella_16,2017A&A...600A..72F} are used to infer the depth of the gaps in gas surface densities, the thermal structure needs to be computed taking the possible gas and dust thermal de-coupling into account. Additional observations constraining the gas temperature (e.g., via line ratios, as suggested in Section~\ref{sec:line_ratio}) are indeed needed to verify how important this effect is in reality. A detailed study of the dependence of gap depth in CO on planet mass is included in a follow-up study (Facchini et al. in prep.).

Here we focus on three main observational diagnostics: the distance of the sub-mm ($850\,\mu$m) ring from the gap derived from CO images, the ratio between the location of the CO wall outside the gap and sub-mm ring, and the radial width of the gap. In order to derive properties of the gap from the ray-traced CO images, such as gap radius and width, the following method is applied \citep[see][for a similar procedure applied to scattered light maps]{2016MNRAS.459.2790R}. First, the CO radial intensity profile is extracted, as shown in Fig.~\ref{fig:profile_surf_dens_01}. Then, assuming that a gap in the CO emission is observable, an analytical fit for the radial intensity profile well outside the gap is derived. In particular, a linear fit for the logarithm of the radius dependent line intensity $I$ is obtained via the relation $\log{I}=a+b(R/{\rm AU})$. In this work, we performed the fit  between $35$ and $60\,$AU, i.e., well outside the region where the intensity profile shows a dip. To highlight the gap, we then used the fitted intensity profile to normalize the entire radial intensity profile. The depth of the gap was estimated as the minimum of the normalized intensity profile $\bar{I}$. The width of the gap was derived as the distance between the two radii, where the normalized intensity profile becomes lower than $(1-0.66\times(1-\bar{I}_{\rm min}))$ \citep{2016MNRAS.459.2790R}, where the outer of these two radii is defined as the CO wall ($R_{\rm CO}$). We defined the gap radius ($R_{\rm gap}$) as the midpoint between these two radii. Finally, the radius of the sub-mm ring ($R_{\rm mm}$) was defined as the radius of maximum intensity in the radial continuum intensity profile outside the gap radius. The distance of this ring from the gap is $R_{\rm mm}-R_{\rm gap}$. A visualization of these three radial parameters can be found in Fig.~\ref{fig:maps_alpha3}.

\begin{figure}
\center
\includegraphics[width=\columnwidth]{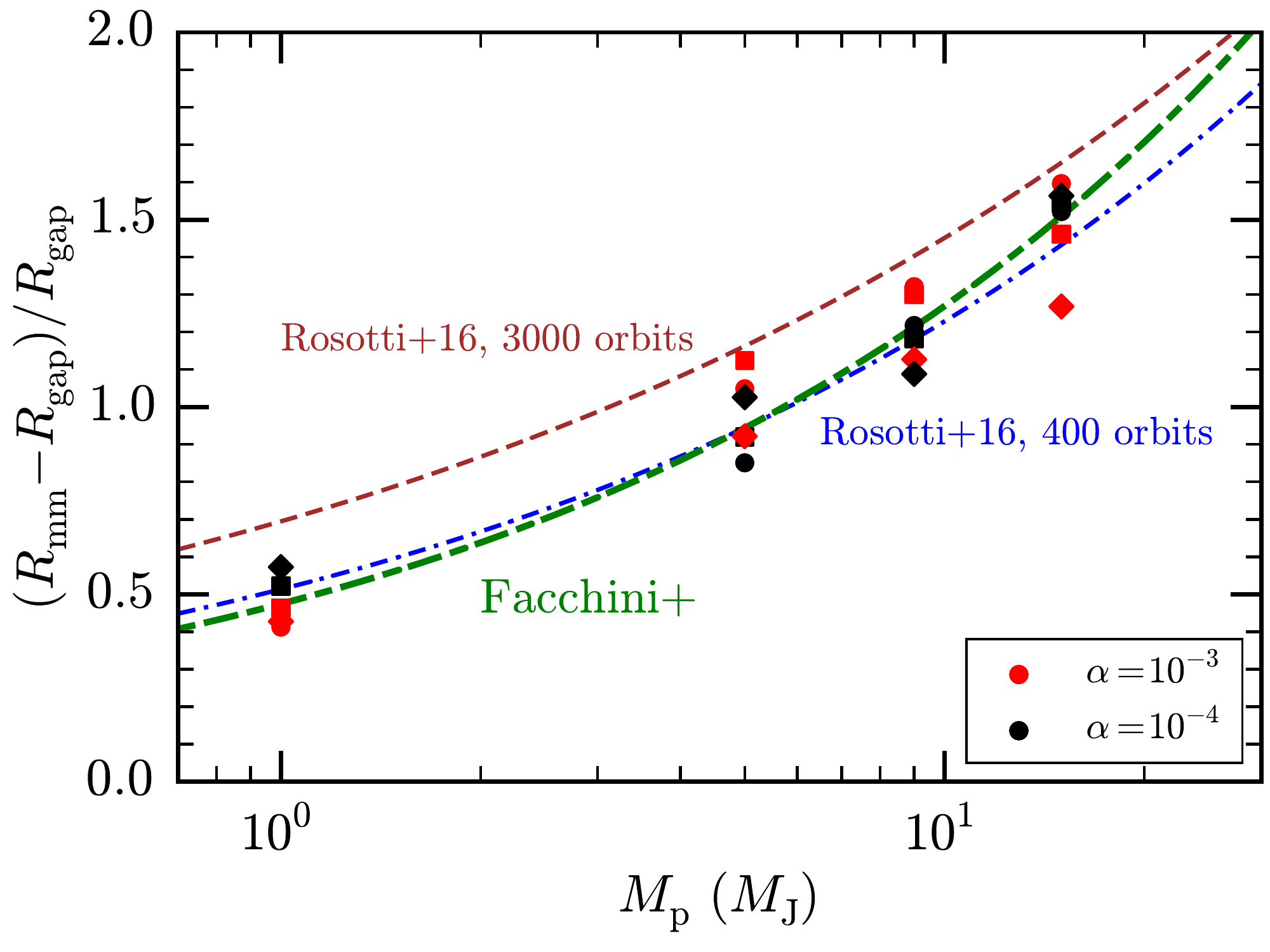}
\caption{Distance of the sub-mm ring $R_{\rm mm}$ from the gap radius $R_{\rm gap}$ as a function of planet mass for all the simulations. The gap radius is estimated from the $^{13}$CO $J$=3-2 emission, whereas the ring radius is obtained from the maximum in sub-mm ($850\,\mu$m) continuum emission outside the gap radius. The ring distance is defined as the ring radius minus the gap radius. Circles, squares, and diamonds indicate quantities retrieved from images convolved with a $0.03\arcsec$, $0.08\arcsec,$ and $0.15\arcsec$ beam, respectively, for sources located at a distance of $150\,$pc. The dashed green line shows a linear fit obtained for this relation in log-log space. The blue dashed-dotted line and brown dashed line indicate the fitting formula by \citet{2016MNRAS.459.2790R} obtained from hydro-simulations with 400 and 3000 planet orbits, respectively, where the gap location was inferred from synthetic scattered light maps.
}
\label{fig:r_mm}
\end{figure}

\subsubsection{Distance of the sub-mm ring from the gap radius}

The distance of the sub-mm ring from the gap radius (normalized to the gap radius itself) as a function of planet mass is presented in Fig.~\ref{fig:r_mm}. The plot shows a clear trend in which the distance of the ring increases with planet mass. The plot contains quantities derived for all eight simulations, and where the analyzed images were produced with three different beam sizes, specifically $0.03\arcsec$, $0.08\arcsec$, and $0.15\arcsec$. The plot shows planet radii as derived from $^{13}$CO. The same result holds for planet radii obtained from $^{12}$CO or C$^{18}$O, since for these three CO isotopologs, and for the beam sizes considered here (up to $0.15\arcsec$), the gap radius is always recovered with an accuracy $<1.5\,$AU. The very small scatter of the ring distance for a given planet mass shows that this diagnostic does not depend on the disk viscosity, at least for the range of viscosities and planet masses analyzed here. Moreover, this diagnostic does not require the gap to be radially resolved; the only requirement is that for a given gap width and depth, the angular resolution is high enough to pick the presence of a gap. This is enforced by the rather small scatter with beam sizes. A linear fitting formula in log-log space can easily be retrieved, using all the points shown in Fig.~\ref{fig:r_mm}, leading to the simple relation

\begin{equation}
\log_{10}{ \left(\frac{R_{\rm mm} - R_{\rm gap}}{R_{\rm gap}}\right)} = A + B\log_{10}{\left( \frac{M_{\rm p}}{M_{\rm J}} \right)},
\end{equation}
where $A=-0.324$ and $B=0.428$. Interestingly, the relation between ring distance and planet mass is also well resembled by the semi-analytical fits derived by \citet[][see their fig.~17]{2016MNRAS.459.2790R}, shown in Fig.~\ref{fig:r_mm}, where the gap location is derived from scattered light images. In the simulations by Rosotti et al., the range of planet masses is rather different from the simulation analyzed here (between $8$ and $120\,M_\oplus$), indicating that this relation holds for a rather large range of planet masses. Moreover, \citet{2016MNRAS.459.2790R} varied the aspect ratio of the disk at the planet location by a factor of 2 above and below the $0.05$ value used in this paper, finding that the scatter is small. Extending this result to our planet mass range, we would expect this result to hold in our case as well. However this is still to be tested. Finally, the trend shown in Fig.~\ref{fig:r_mm} is better described by the semi-analytical fit for hydrodynamical simulations of $400$ orbits in \citet{2016MNRAS.459.2790R}, as opposed to the $3000$ orbits fit. Since our simulations run for $1000$ planet orbits, this may indicate that the gas surface densities are not completely in (but very close to) steady state, or that the rather different treatment of the dust dynamics, together with the different planetary mass regime, can lead to slightly different results.

\subsubsection{Ratio of sub-mm radius and CO wall}

\begin{figure}
\center
\includegraphics[width=\columnwidth]{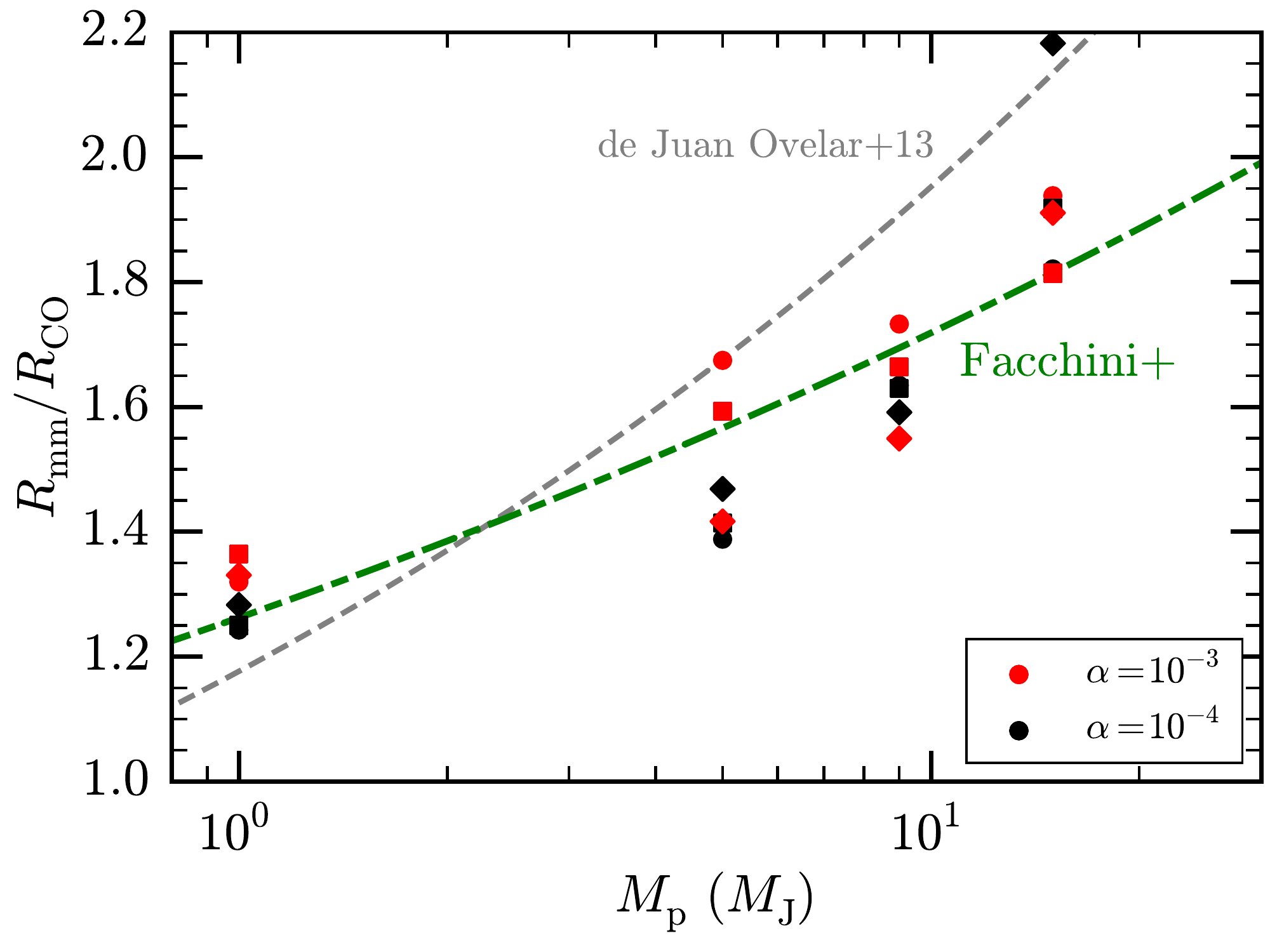}
\caption{Ratio between the sub-mm ring $R_{\rm mm}$ and  radius of the CO wall $R_{\rm CO}$. For this particular plot, the $^{13}$CO $J$=3-2 line is used. As in Fig.~\ref{fig:r_mm}, circles, squares, and diamonds indicate quantities retrieved from images convolved with a $0.03\arcsec$, $0.08\arcsec$, and $0.15\arcsec$ beam, respectively, for sources located at a distance of $150\,$pc. The dashed green line shows a linear fit obtained for this relation in log-log space. The gray dashed line indicates the fit by \citet{2013A&A...560A.111D} for the sub-mm radius/scattered light wall, obtained from a subsample of the same hydrodynamical simulations used in this paper.}
\label{fig:r_mm_djo}
\end{figure}

A second diagnostic is to compare the sub-mm radius with the CO radius, defined as the location outside the gap where CO reaches $0.33$ times the flux difference between the bottom of the gap and the maximum flux outside the gap. This can be of particular interest for transition disks that show a CO cavity rather than a\  CO gap, since the definition of CO radius still holds in the former case.The dependency of $R_{\rm mm}/R_{\rm CO}$ is shown in Fig.~\ref{fig:r_mm_djo}, where the $^{13}$CO $J$=3-2 line is used. There is a clear trend in which the ratio increases with planet mass. However, when compared to the previous diagnostic, the vertical scatter is much larger. This is because the gap needs to be resolved  to obtain a good estimate of where the CO wall is located. At $0.08\arcsec$ resolution, the gap starts to be unresolved for $M_{\rm p}\gtrsim5M_{\rm J}$. As for the previous diagnostic, a linear fit in log-log space can be found for the dependency of $R_{\rm mm}/R_{\rm CO}$ on planet mass,

\begin{equation}
\log_{10}{ \left(\frac{R_{\rm mm}}{R_{\rm CO}}\right)} = \tilde{A} + \tilde{B}\log_{10}{\left( \frac{M_{\rm p}}{M_{\rm J}} \right)},
\end{equation}
where $\tilde{A}=0.101$ and $\tilde{B}=0.134$. In the fitting, only points with angular resolution $\leq0.08\arcsec$ were used. The analytical formula is similar to that provided by \citet[][see Fig.~\ref{fig:r_mm_djo}]{2013A&A...560A.111D} for a subsample of the hydrodynamical simulations explored in this paper, where instead of $R_{\rm CO}$ they used a similar definition for the scattered light intensity profile. Their analytical fit shows a steeper dependence on planet mass, indicating that the wall in scattered light may extend at slightly larger radii than the relative peak in the CO emission.

\begin{figure*}
\begin{center}
\includegraphics[width=0.32\textwidth]{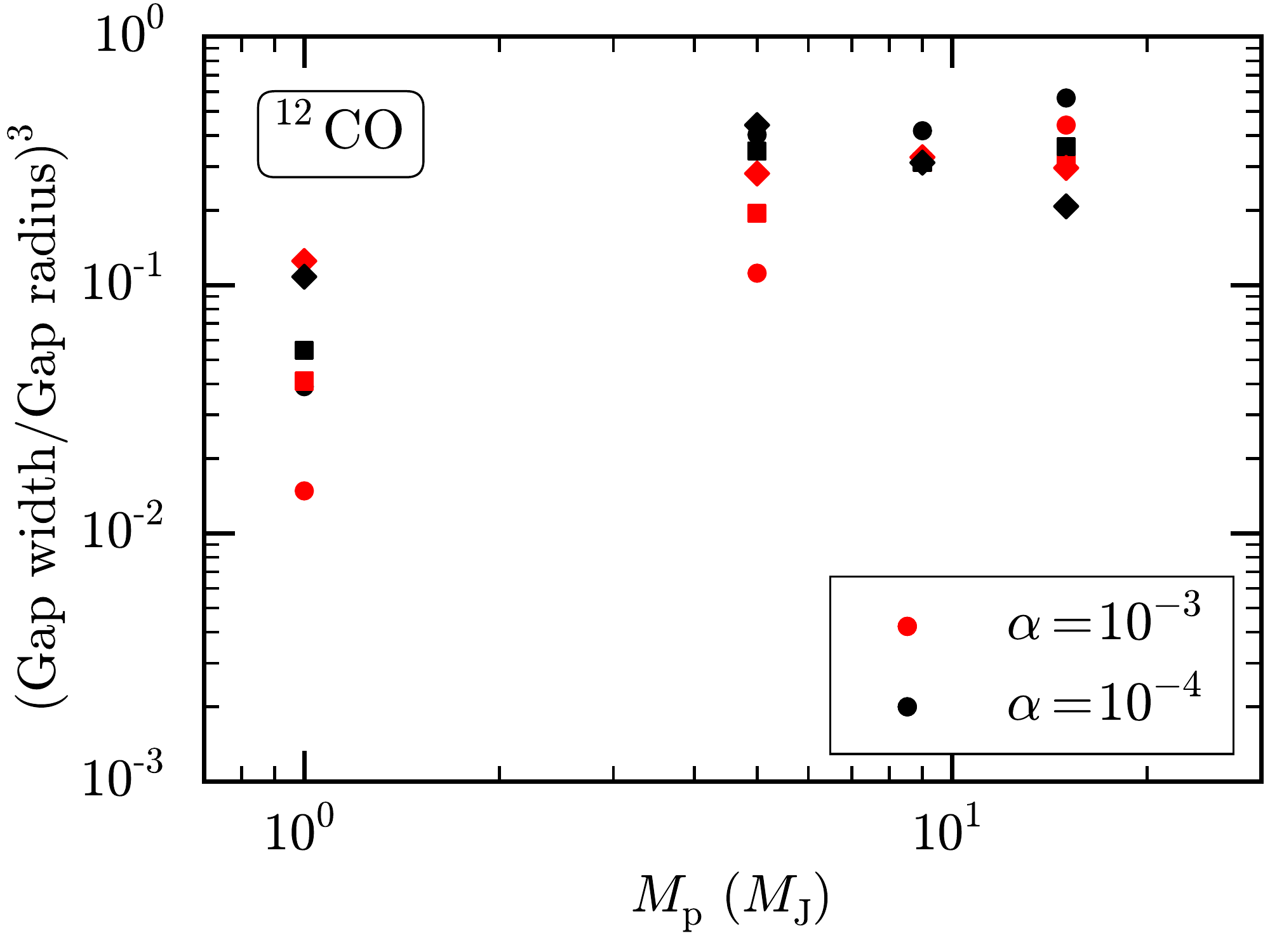}
\includegraphics[width=0.32\textwidth]{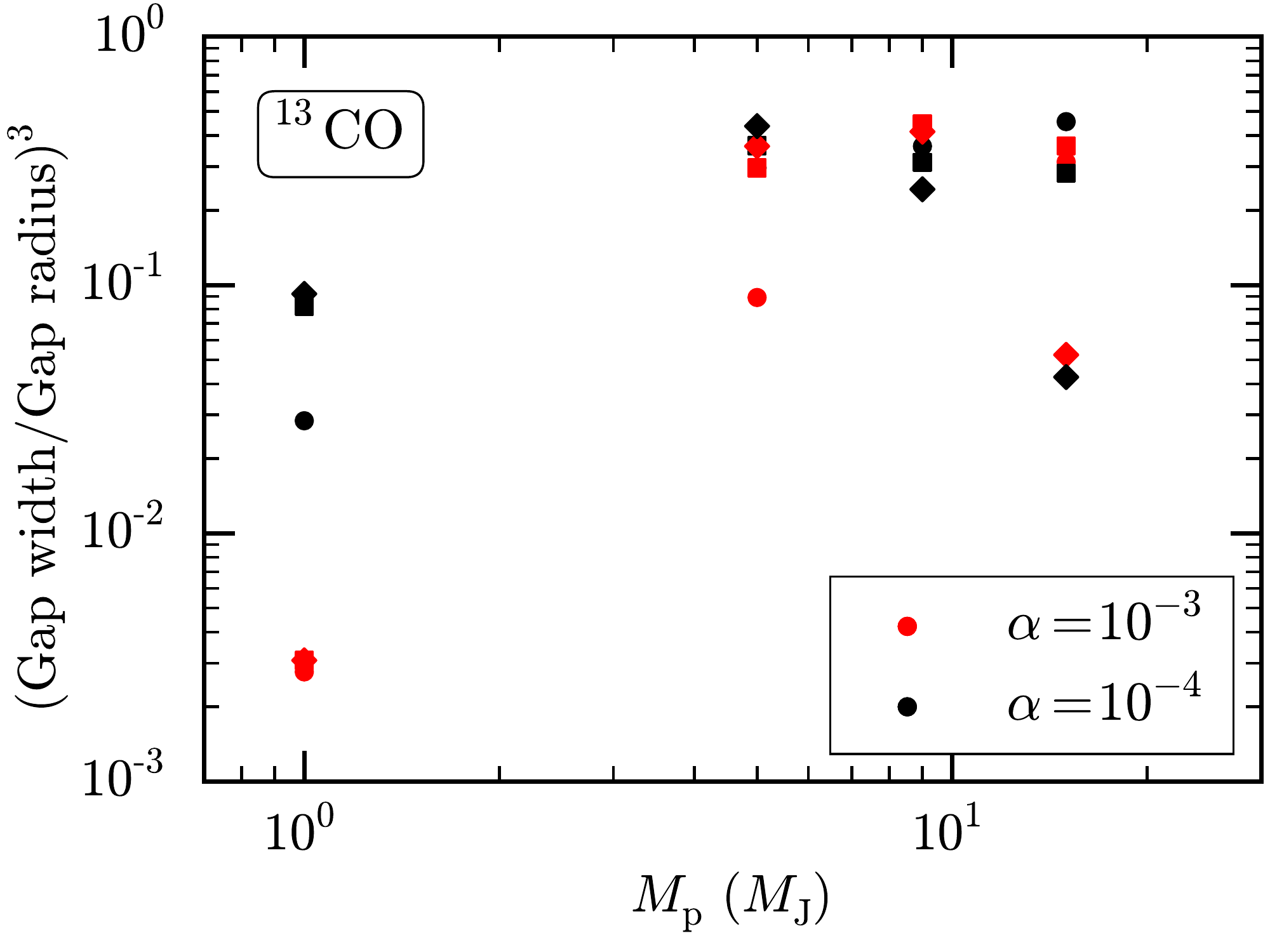}
\includegraphics[width=0.32\textwidth]{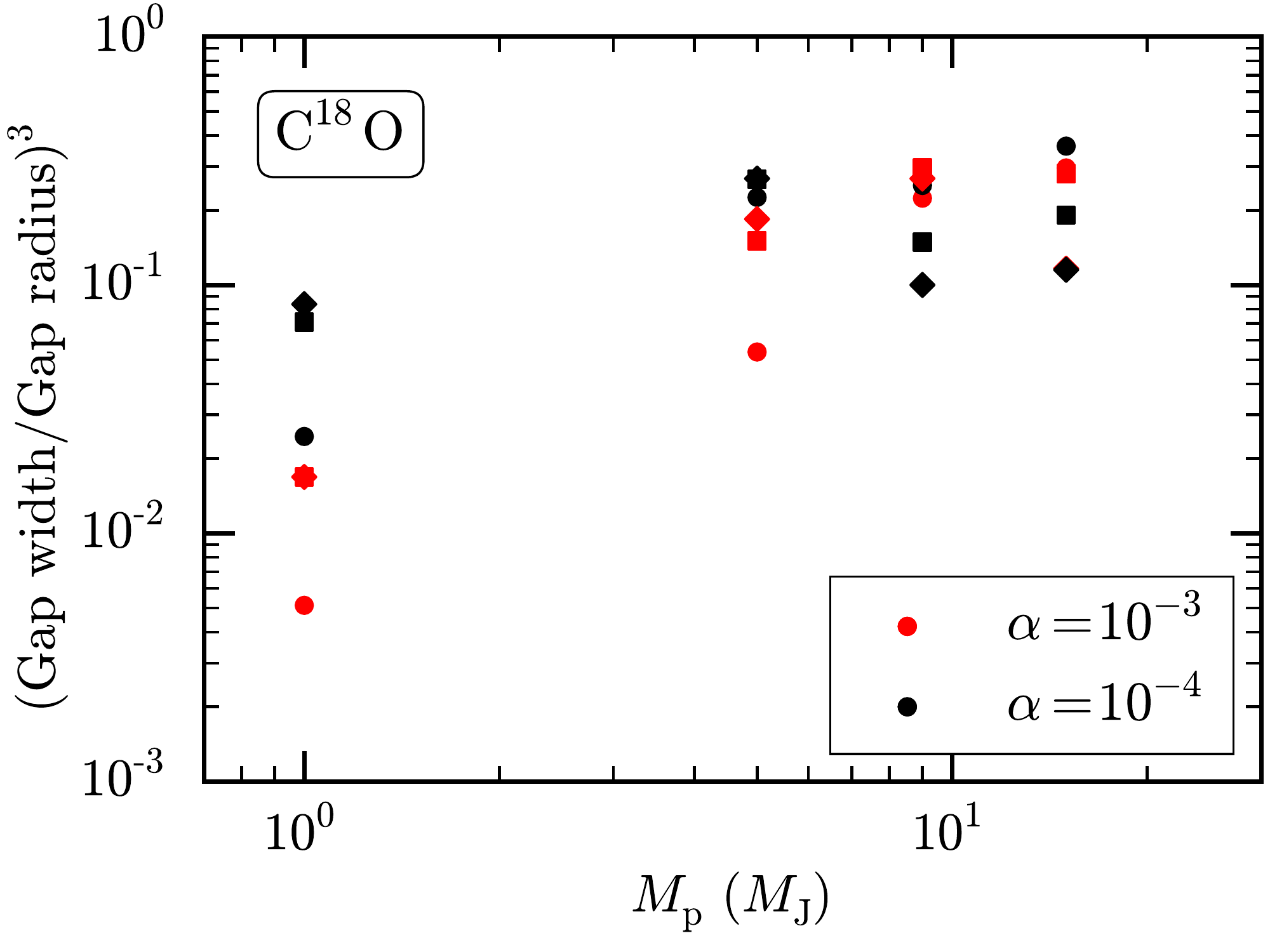}\\
\end{center}
\caption{Gap width normalized to the gap radius $R_{\rm gap}$ as a function of planet mass. From left to right: gap width and radius determined from $^{12}$CO, $^{13}$CO, and C$^{18}$O intensity profiles, respectively. The third power of the gap width is used to emphasize the dependency. As in Fig.~\ref{fig:r_mm}, circles, squares, and diamonds indicate quantities retrieved from images convolved with a $0.03\arcsec$, $0.08\arcsec$, and $0.15\arcsec$ beam, respectively, for sources located at a distance of $150\,$pc. 
}
\label{fig:delta_r}
\end{figure*}

\subsubsection{CO gap width}

The third diagnostic to infer the planet mass, i.e., the gap width derived from CO images, is less straightforward. The dependence of the gap width on planet mass is shown in Fig.~\ref{fig:delta_r}, where the gap width has been estimated from the three main CO isotopologs, $^{12}$CO, $^{13}$CO, and C$^{18}$O. In general, there is a clear trend of gap width increasing with planet mass, as we would expect. The high mass end of the plots ($M_{\rm p}\geq9M_{\rm J}$) is consistent for all three CO isotopologs, where the dependence of the gap width on mass saturates, as shown so in the gas surface density profiles in Fig.~\ref{fig:profile_surf_dens_01}. At these large gap widths, the vertical scatter for a given planet mass is low, indicating that the gap size is comparable to the largest beam size used in the convolution, and that the gap width saturates at roughly the same values for both values of disk viscosity. For lower masses, however, the picture is more complicated. First, there is significant difference among the widths derived by the three CO isotopologs owing to optical depth  and temperature effects. Second, the gap width does depend on viscosity, as clearly shown in the $^{13}$CO and C$^{18}$O cases. Third, the estimate of the gap width depends on the angular resolution, in particular for the $1M_{\rm J}$ cases, where the gap is unresolved for beam sizes $\geq0.08\arcsec$. With beam sizes larger than the gap width, the latter are systematically overestimated, with derived width increasing with beam size.

Given the potentially large scatter in the estimated gap width, we consider the relation between the distance of the (sub-)mm ring from the gap seen in CO as more robust, since this relation:
\begin{itemize}
\item is not affected by thermal variations within the gap;\item does not depend significantly on viscosity;
\item does not require a very high-angular resolution (at least for the massive planets considered in this paper), i.e., it does not require the gap to be radially resolved;
\item shows good consistency using different CO isotopologs as gap tracers;
\item \citet{2016MNRAS.459.2790R} have shown that this same relation has little dependency on the aspect ratio for planet masses $<1M_{\rm J}$, and thus we would expect this to hold true for higher mass planets, such as those considered in this paper;
\item a good semi-analytical prescription seems to well describe the relation both for massive planets (this paper) and lighter planets \citep{2016MNRAS.459.2790R}; further studies are needed to confirm that this relation holds true for planet masses $<1M_{\rm J}$ when CO is used to trace the gap, instead of scattered light.
\end{itemize}
The relation between the (sub-)mm radius and the CO wall is also robust within the parameter range explored in this paper, even though it requires very high angular resolution to be used safely. In summary,
\begin{itemize}
\item it is not significantly affected by thermal variations within the gap;
\item it does not depend significantly on viscosity;
\item it does show good consistency using different CO isotopologs as gap tracers; and
\item it may be used to infer planet masses in transition disks that show a CO cavity, where no gap is observed.
\end{itemize}
The dependency on disk mass might also be important. Follow-up dedicated studies are needed to address this possibility.

\section{Conclusions}
\label{sec:concl}

This work is the first study to compute the gas temperature and chemical abundances in protoplanetary disks hosting one single massive planet ($M_{\rm p}\geq1\,M_{\rm J}$), where the gas surface densities and the radial grain size distributions were taken from existing hydrodynamical and dust evolution simulations. Synthetic images of both (sub-)mm continuum and CO rotational lines for different disk and planet properties are provided. The (sub-)mm continuum emission is typical of transition disks. We demonstrated how simultaneous observations of mm continuum and CO lines provide strong constraints on the presence and mass of embedded planets. The main conclusions can be listed as follows:

\begin{enumerate}
\item The distance between the location of the gap seen in CO and the (sub-)mm continuum ring can be used to infer the planet mass. This diagnostic shows very small scatter, and it seems independent of disk viscosity and angular resolution, provided that the gap is visible (but not necessarily resolved) in the intensity profile.
\item The ratio of the location of the (sub-)mm ring and the CO wall shows a clear dependency on planet mass. However, high angular resolution observations are needed to use this diagnostic to infer planet masses. This relation can be used when no CO gaps are observed, but only CO cavities.
\item It is difficult to estimate the planet mass using only the gap width of the CO lines. However, there is a general trend of the gap width increasing with planet mass. This relation does depend on viscosity and angular resolution, if the gap width is not resolved.
\item CO emission maps can provide the same or better constraints on embedded planets as scattered light maps when a gap is observed in CO emission, since gaps in CO and in scattered light are expected to be co-spatial. The advantage of using rotational molecular lines is that they are obtained simultaneously with the continuum with a single instrument and that scattered light observations taken from the ground are limited to bright sources.
\item The gas in the gap is colder than the dust component as a result of low thermal coupling between gas and dust.
\item CO line ratios can be used to track thermal gaps and distinguish such gaps from surface density gaps. The best line combination is with the $J$=6-5 and $J$=2-1 lines because of the energy difference between the two transitions.
\item Obtaining gas surface densities across sub-mm continuum gaps from CO observations without modeling the gas temperature structure properly can lead to large overestimates of the planet masses. 
\item No CO cavities are seen in these models with only one massive planet; only CO gaps are found, in contrast with observations.
\end{enumerate}

Further modeling is required to predict molecular line emission from disks hosting less massive planets. Upcoming ALMA and scattered light observations are going to further constrain the models presented in this paper.

\begin{acknowledgements}
We thank the anonymous referee, whose comments helped improving the clarity of the paper. We are grateful to Giovanni Rosotti for useful discussions and for early comments on the manuscript. Astrochemistry in Leiden is supported by the European Union A-ERC grant 291141 CHEMPLAN, by the Netherlands Research School for Astronomy (NOVA) and by a Royal Netherlands Academy of Arts and Sciences (KNAW) professor prize. P.P. acknowledges support by NASA through Hubble Fellowship grant HST-HF2-51380.001-A awarded by the Space Telescope Science Institute, which is operated by the Association of Universities for Research in Astronomy, Inc., for NASA, under contract NAS 5-26555. All the figures were generated with the \textsc{python}-based package \textsc{matplotlib} \citep{2007CSE.....9...90H}.
\end{acknowledgements}


\bibliographystyle{aa}
\bibliography{references}

\begin{appendix}

\section{Dust and gas temperatures}

\label{app:temp}

\begin{figure*}
\begin{center}
\includegraphics[width=\textwidth]{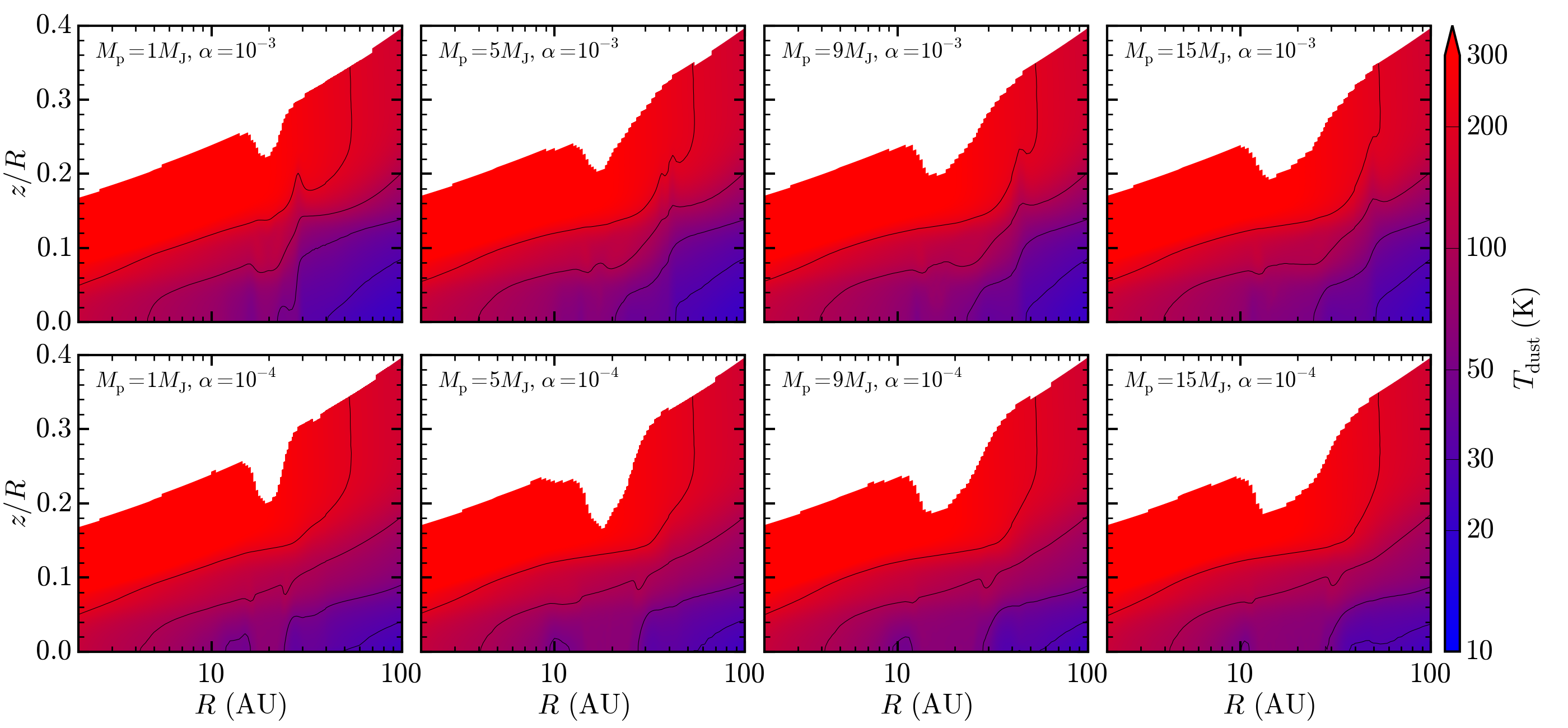}
\end{center}
\caption{Dust temperatures for all models.}
\label{fig:tdust}
\end{figure*}

\begin{figure*}
\begin{center}
\includegraphics[width=\textwidth]{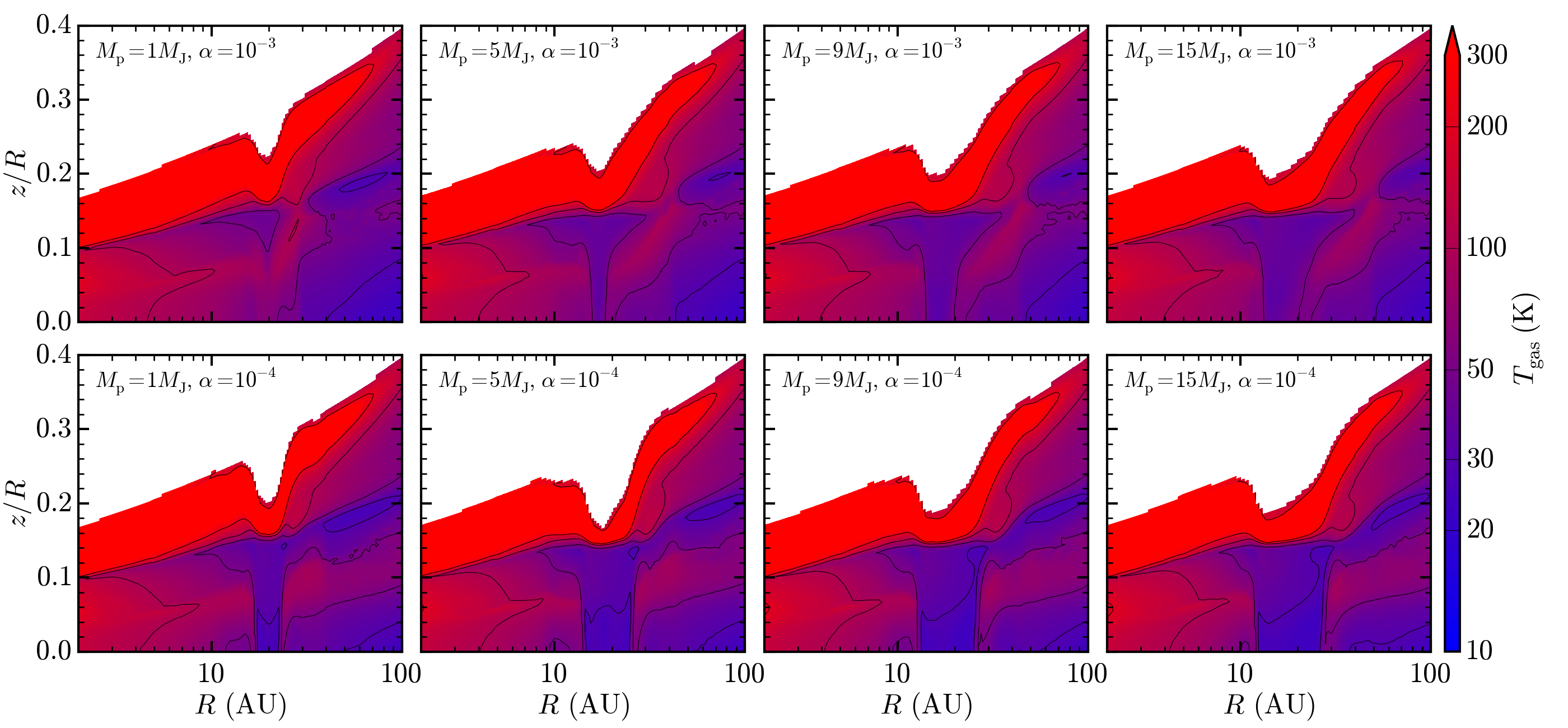}
\end{center}
\caption{Gas temperatures for all models.}
\label{fig:tgas}
\end{figure*}

\begin{figure*}
\begin{center}
\includegraphics[width=\textwidth]{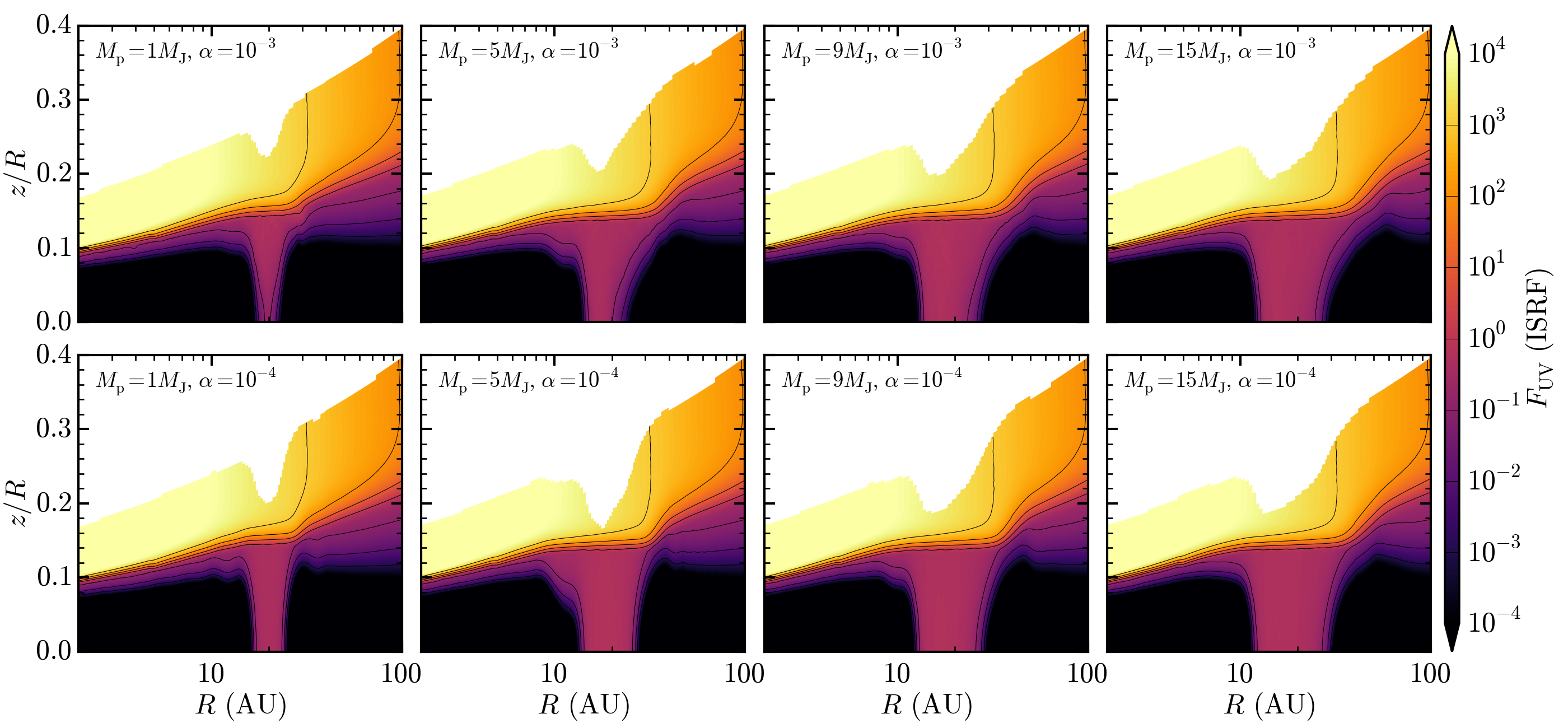}
\end{center}
\caption{Far-ultraviolet fluxes in $G_0$ units for all models.}
\label{fig:g0}
\end{figure*}

In Fig.~\ref{fig:tdust} and \ref{fig:tgas}, we show the dust and gas temperature of all models. The two quantities are used to produce the temperature ratio reported in Fig.~\ref{fig:temp_ratio}. In particular, Fig.~\ref{fig:tdust} indicates how the dust temperature increases by a few K within the gap opened by the planet. This effect is mostly visible for the $\alpha=10^{-4}$ simulations, where the dust surface density in the proximity of the perturbing planet is very low. Conversely, Fig.~\ref{fig:tgas} shows that the gas temperature becomes colder within the gaps, as explained in more detail in Section~\ref{sec:dens_temp}. The significance of this effect increases with planet mass and decreases with turbulence. Fig.~\ref{fig:g0} shows the local FUV fluxes. The FUV photons can partly penetrate within the gap carved by the planet.

\section{Abundances}
\label{app:abu}

\begin{figure*}
\begin{center}
\includegraphics[width=0.32\textwidth]{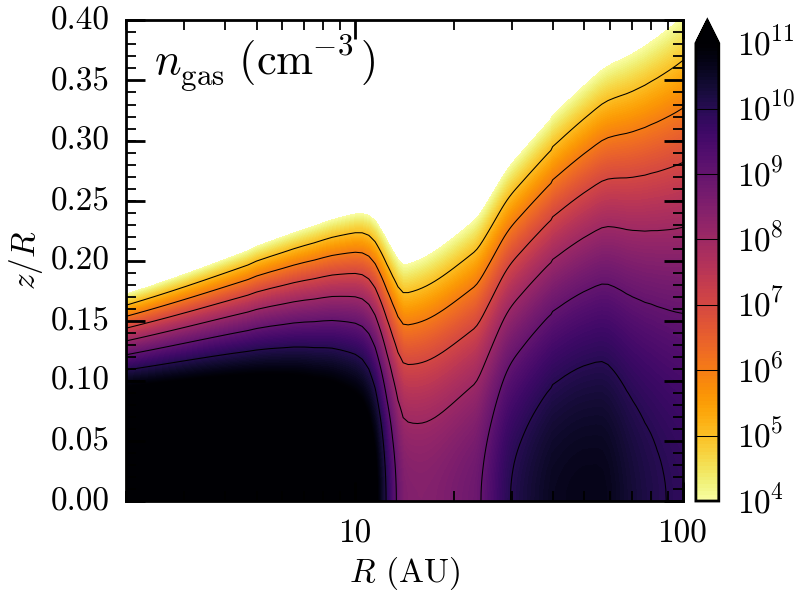}
\includegraphics[width=0.32\textwidth]{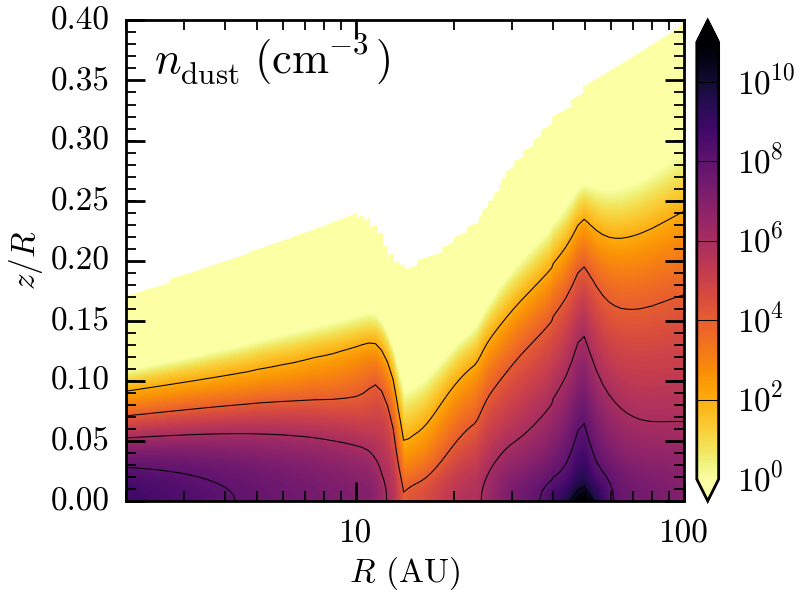}
\includegraphics[width=0.32\textwidth]{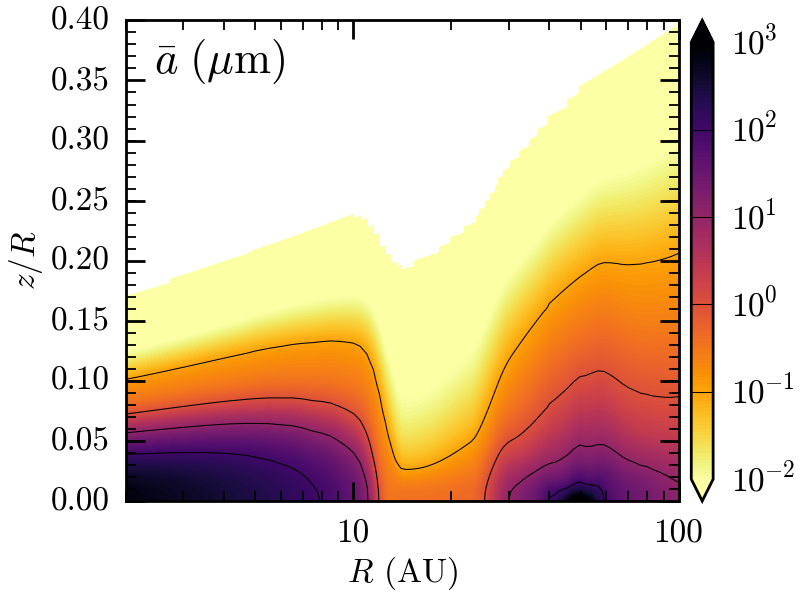}\\
\includegraphics[width=0.32\textwidth]{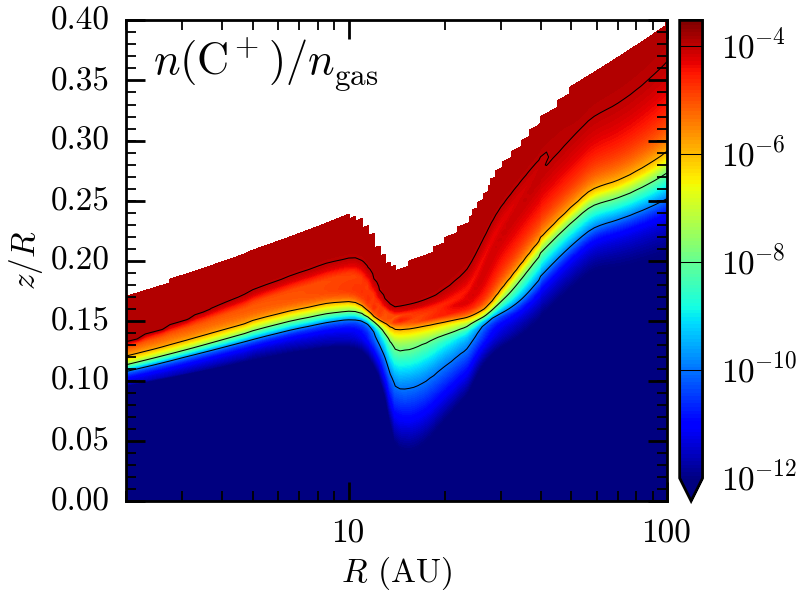}
\includegraphics[width=0.32\textwidth]{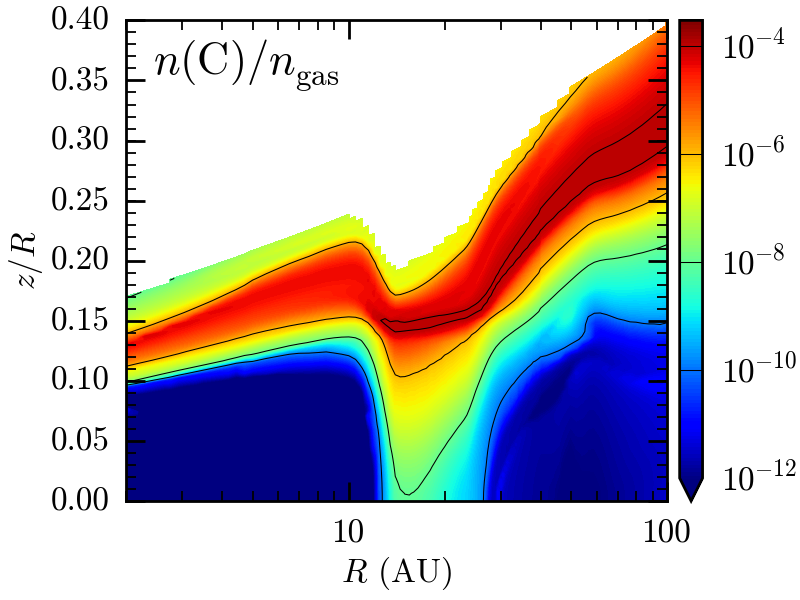}
\includegraphics[width=0.32\textwidth]{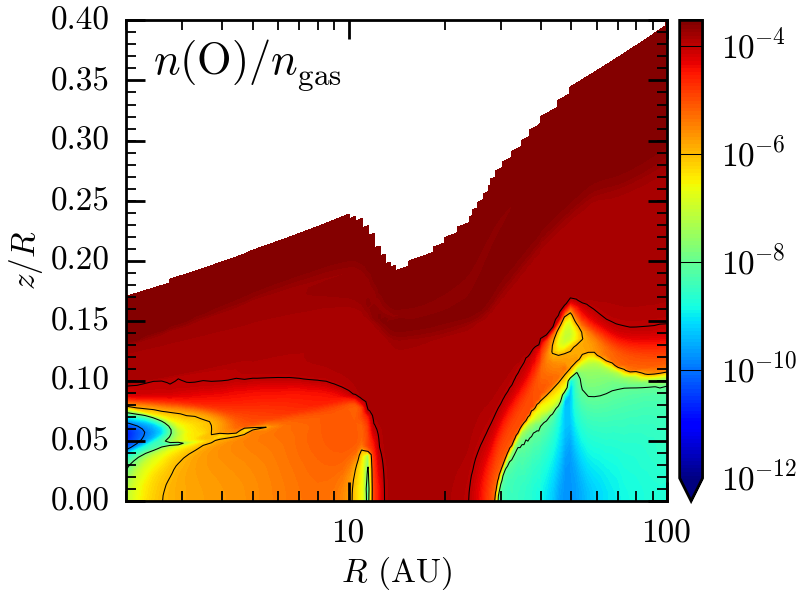}\\
\includegraphics[width=0.32\textwidth]{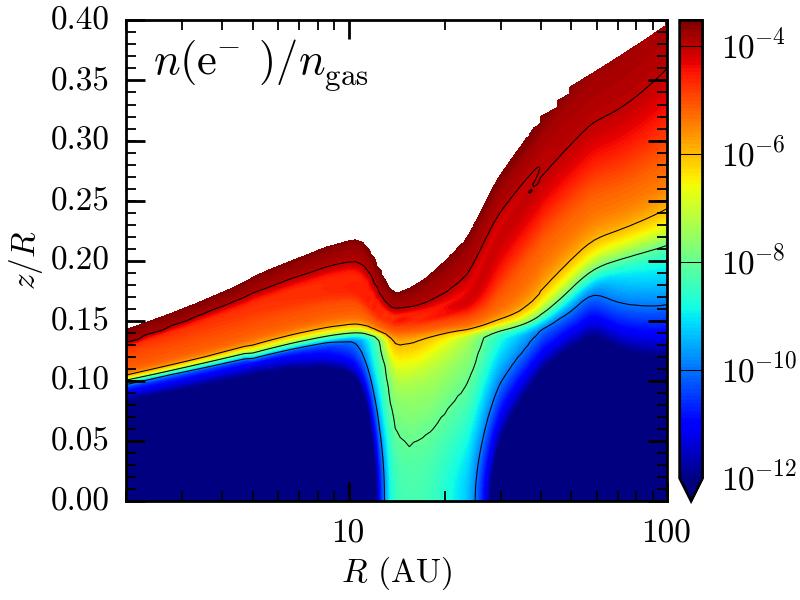}
\includegraphics[width=0.32\textwidth]{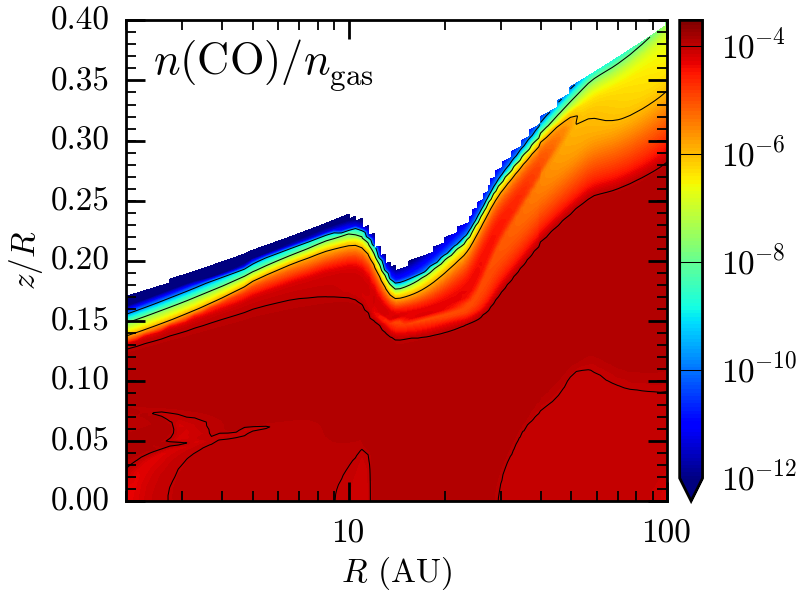}
\includegraphics[width=0.32\textwidth]{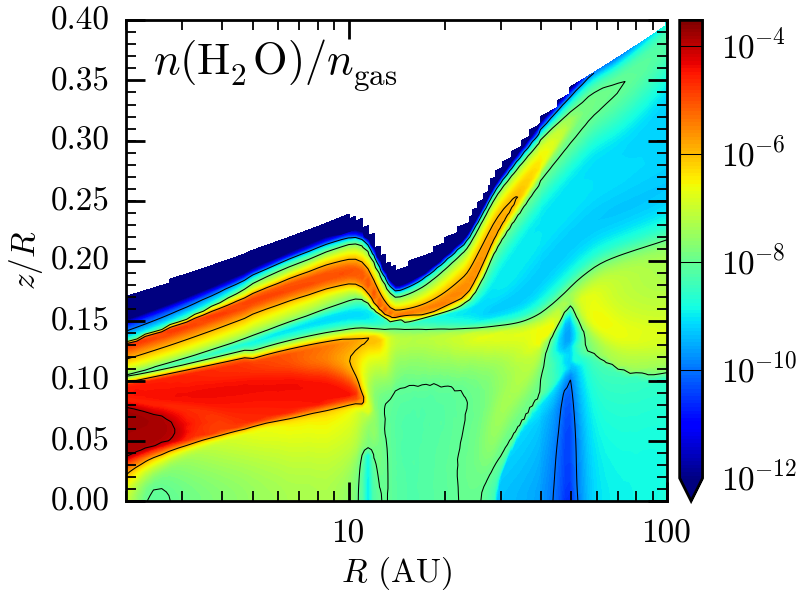}\\
\includegraphics[width=0.32\textwidth]{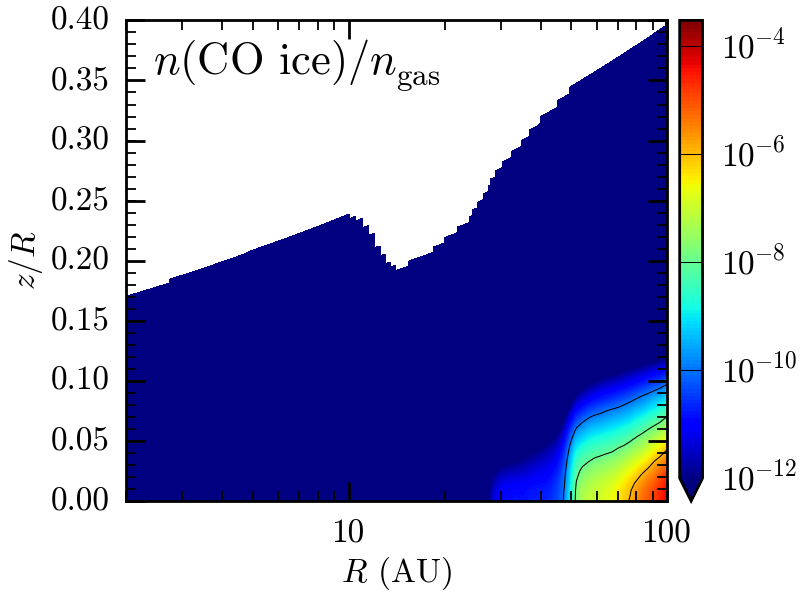}
\includegraphics[width=0.32\textwidth]{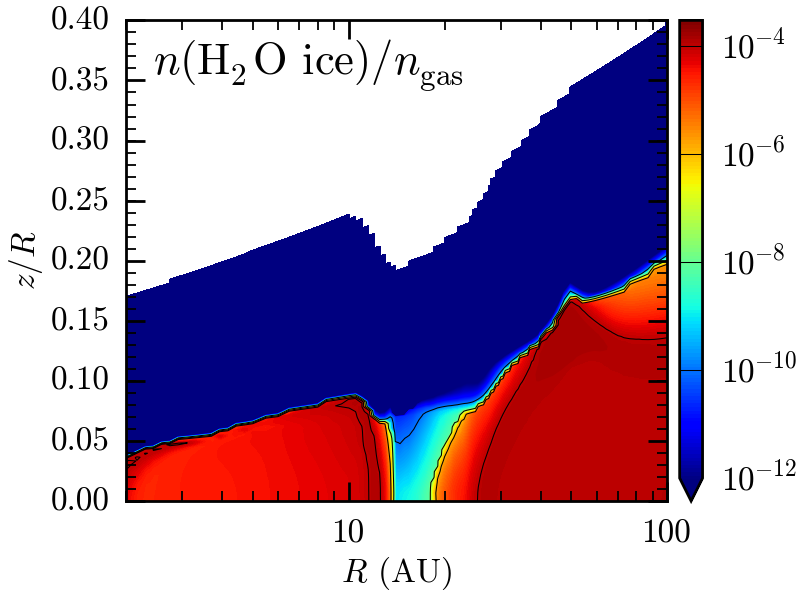}
\includegraphics[width=0.32\textwidth]{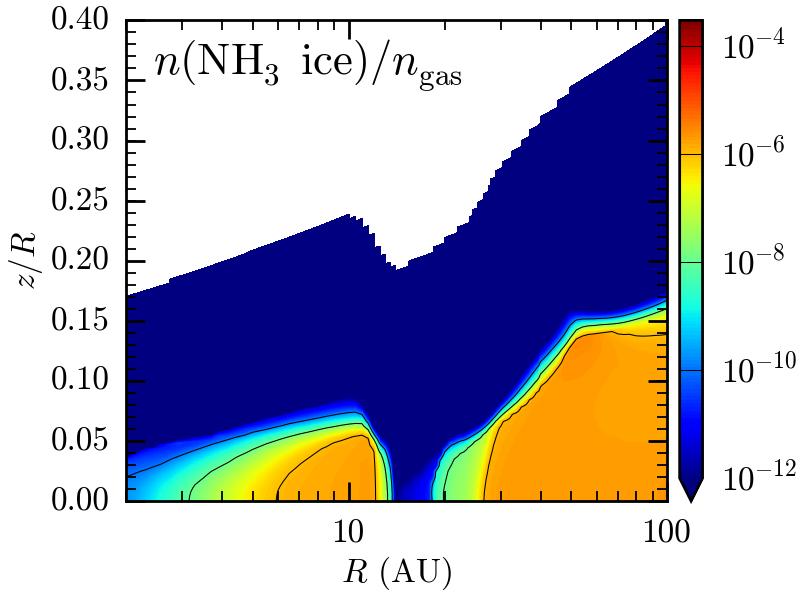}\\
\end{center}
\caption{Density structure, average grain size, and gas, and ice abundances for the $\alpha=10^{-3}$, $M_{\rm p}=15\,M_{\rm J}$ model.
}
\label{fig:abu}
\end{figure*}

Fig.~\ref{fig:abu} shows the density structure and abundances of a few key volatile and ice species for the $\alpha=10^{-3}$, $M_{\rm p}=15\,M_{\rm J}$ model.

\section{Emission maps of the $\alpha=10^{-4}$ models}
\label{app:maps_alpha4}

\begin{figure*}
\begin{center}
\includegraphics[width=.85\textwidth]{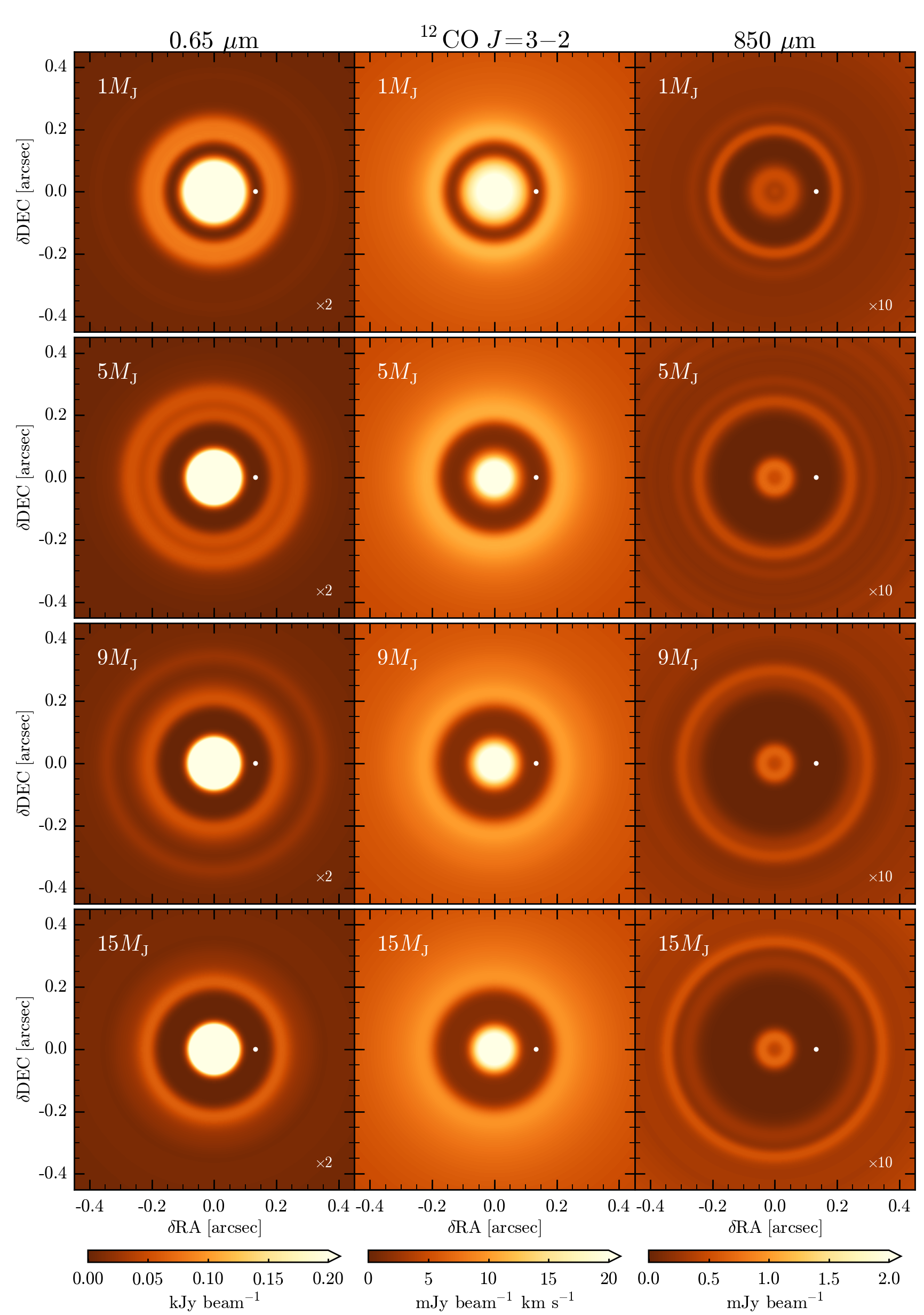}
\end{center}
\caption{As in Fig.~\ref{fig:maps_alpha3}, for the $\alpha=10^{-4}$ simulations. The emission maps in scattered light and sub-mm continuum were rescaled up by a factor of $2$ and $10$, respectively.
}
\label{fig:maps_alpha4}
\end{figure*}

Fig.~\ref{fig:maps_alpha4} shows the emission maps in scattered light, $^{12}$CO $J$=3-2 and continuum at $850\,\mu$m of all models with $\alpha=10^{-4}$.

\section{Detectability of CO gaps with ALMA}
\label{app:ALMA}

We performed simulated ALMA observations of all models for the $^{12}$CO $J$=3-2 and $J$=6-5 lines. The source declination was set to $-25\degr$, and the observations were symmetric to transit. Fig.~\ref{fig:line_ratios} shows that an angular resolution $\lesssim0.1\arcsec$ is sufficient in detecting the gap in both lines, thus we chose antenna configurations that deliver such a resolution. In particular, the C43-7 configuration was used for the Band 7 simulated observations of the $^{12}$CO $J$=3-2 line at 345\,GHz, and the C43-5 configuration for the Band 9 observations of the $^{12}$CO $J$=6-5 line at 691\,GHz. The spectral resolution was set to $0.1\,$km\,s$^{-1}$, and the integration time to $1.5\,$h. The simulated observations and consequent continuum subtraction and cleaning were performed with the {\tt simobserve}, {\tt simanalyze,} and {\tt clean} tasks in Common Astronomy Software Applications (CASA) v4.7.2. The synthesized beam is $0.094\arcsec\times0.103\arcsec$ for the Band 9 observations, and $0.067\arcsec\times0.079\arcsec$ for the Band 7 observations. Thermal noise was considered with a system temperature of $269\,$K,  optimal atmospheric conditions for the Band 9 observations ($0.472$\,mm of water vapor column), and good atmospheric conditions for the Band 7 observations ($0.913$\,mm of water vapor column). These settings imply a sensitivity of $31.9\,$mJy in one velocity channel for the $^{12}$CO $J$=6-5 observations, and of $5.13\,$mJy in one velocity channel for the $^{12}$CO $J$=3-2 observations.

\begin{figure*}
\begin{center}
\includegraphics[width=\textwidth]{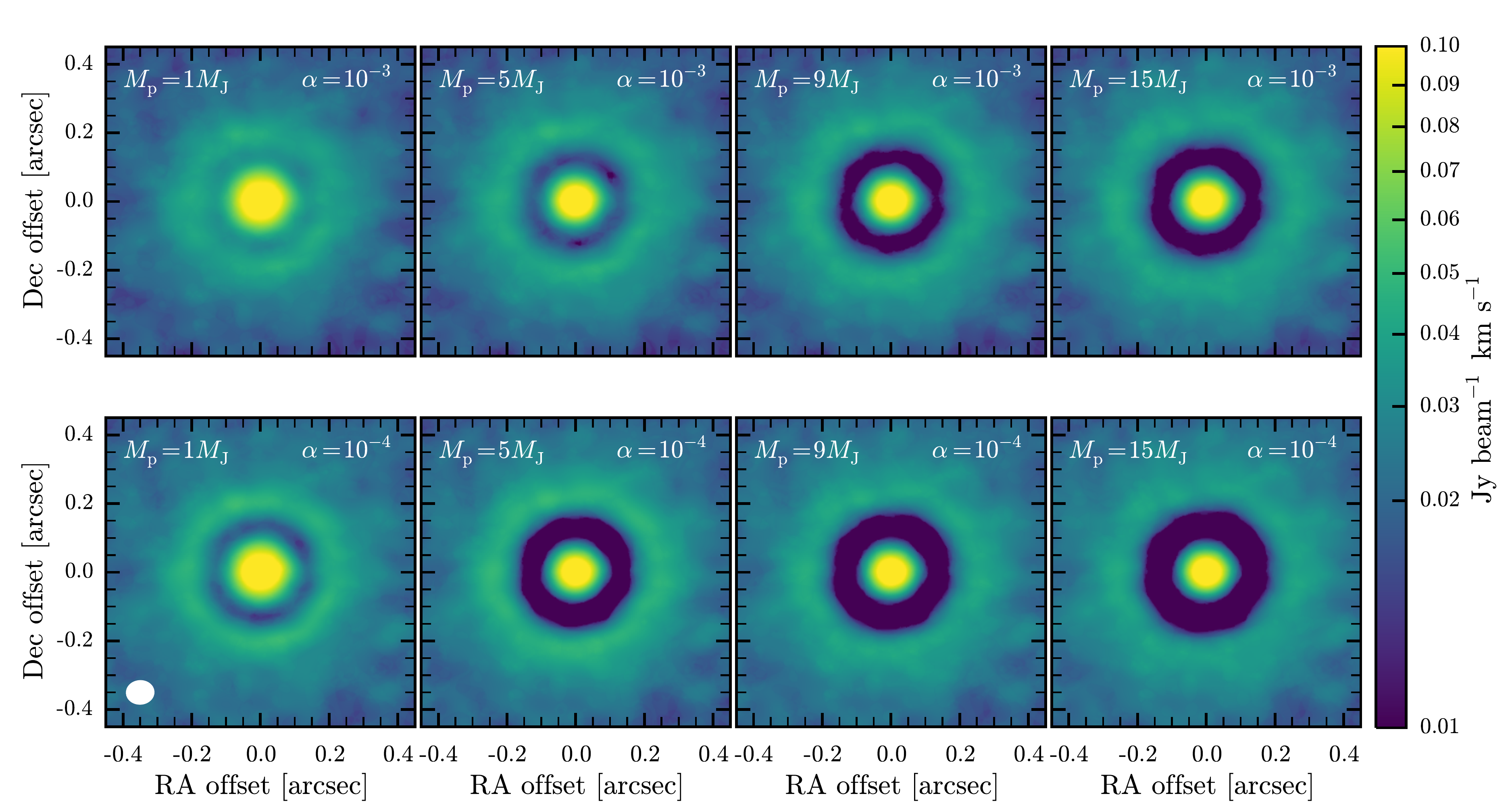}
\end{center}
\caption{ALMA simulated observations of the $^{12}$CO $J$=3-2 line of all models. The distance from the source is 150\,pc. The synthesized beam is shown in the bottom left corner of the figure. The simulated observations were performed in the C43-7 configuration, for an integration time of $1.5$\,h, and a spectral resolution of $0.1\,$km\,s$^{-1}$.
}
\label{fig:alma_12co32}
\end{figure*}

\begin{figure*}
\begin{center}
\includegraphics[width=\textwidth]{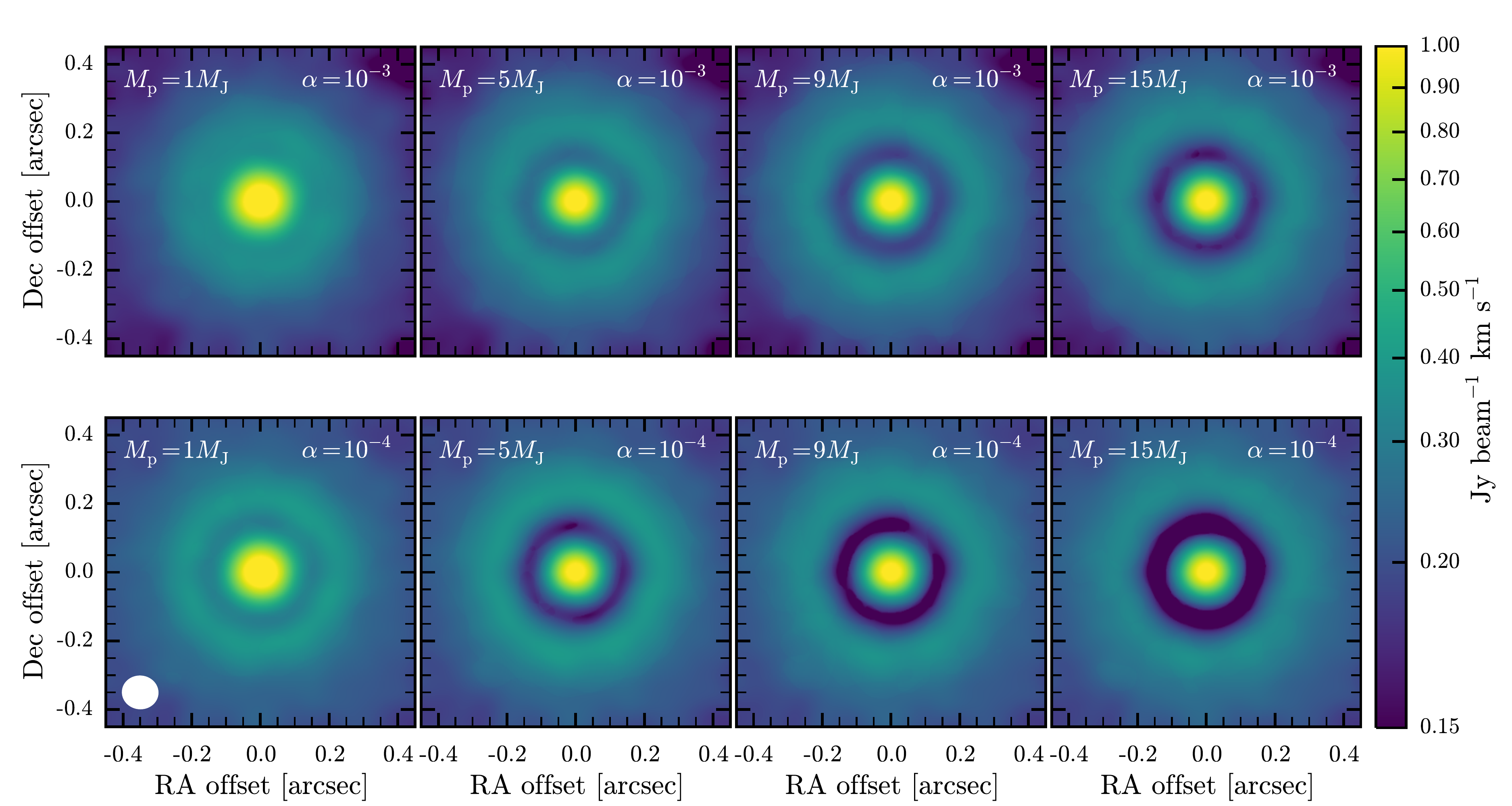}
\end{center}
\caption{ALMA simulated observations of the $^{12}$CO $J$=6-5 line of all models. The assumed distance is 150\,pc. The synthesized beam is shown in the bottom left corner of the figure. The simulated observations were performed in the C43-5 configuration, for an integration time of $1.5$\,h, and a spectral resolution of $0.1\,$km\,s$^{-1}$. 
}
\label{fig:alma_12co65}
\end{figure*}

The ALMA synthetic observations of the $^{12}$CO $J$=3-2 and $J$=6-5 lines are shown in Figs.~\ref{fig:alma_12co32}-\ref{fig:alma_12co65}. The gap in CO is clearly detectable in all models, except for the $M_{\rm p}=1\,M_{\rm J}$ and $\alpha=10^{-3}$ case in the $J$=6-5 line. This is mainly due to the beam size, which smears out the radially thin gap. These simulated observations indicate that a gap carved by a planet with mass $\geq1\,M_{\rm J}$ at $20\,$AU can be detected with high signal to noise in $1.5\,$h integration time, in both ALMA Band 7 and Band 9. We recall that the method used in this paper applies an azimuthal average of the gas density structure computed in the hydrodynamical simulations. For planet masses $<5M_{\rm J}$, the azimuthal asymmetry in the hydrodynamical simulation is small. For higher mass planets, such asymmetry can be more significant (Fig.~\ref{fig:hydro}), and thus real systems hosting such massive planets are expected to have asymmetric molecular line emission, which cannot be reproduced by our models. With the same integration times, the ring in continuum can be characterized with high signal to noise, assuming a typical bandwidth of $7.5\,$GHz. As noted in the introduction, gas cavities and gaps have already been seen by ALMA.

Simulated SPHERE observations of the models have been performed by \citet{2016MNRAS.459L..85D}. The angular resolution attained by typical SPHERE observations is $\sim0.04\arcsec$. The gaps produced by our models can be observed in $\sim1\,$h integration time on VLT, where the exact value depends on a number of factors, in particular the luminosity of the central star in the NIR.

\end{appendix}

\end{document}